\documentclass[aps,prd,
nofootinbib, twoside,
10pt
]
{revtex4}

\usepackage{graphicx}
\usepackage{amssymb, amsmath}
\usepackage[unicode]{hyperref}

\usepackage{epsf}

\usepackage{color}
\usepackage{ulem}

\def\la{\; \raise0.3ex\hbox{$<$\kern-0.75em\raise-1.1ex\hbox{$\sim$}}\;}
\def\ga{\;  \raise0.3ex\hbox{$>$\kern-0.75em\raise-1.1ex\hbox{$\sim$}}\;}
\newcommand{\aap}{Astron.\ Astrophys.}
\newcommand{\mnras}{Mon.\ Not.\ R.\ Astron.\ Soc.}
\newcommand{\mnrasl}{Mon.\ Not.\ R.\ Astron.\ Soc.: Lett.}
\newcommand{\physrep}{Phys.\ Rep.}
\newcommand{\apjl}{Astrophys.\ J.\ Lett.}
\newcommand{\apjs}{Astrophys.\ J.\ Suppl.\ Ser.}
\newcommand{\araa}{Ann.\ Rev.\ Astron.\ Astrophys.}
\newcommand{\apss}{Astrophys.\ Space Sci.}

\begin{document}

\title{
Explaining observations of rapidly rotating neutron stars in LMXBs
}

\author{Mikhail E. Gusakov$^{1,2}$}
\author{Andrey I. Chugunov$^1$}
\author{Elena M. Kantor$^1$}
\affiliation{$^1$ Ioffe Institute, Polytekhnicheskaya 26,
194021 St.-Petersburg, Russia
\\
$^2$St.-Petersburg State Polytechnical University,
Polytekhnicheskaya 29, 195251 St.-Petersburg, Russia }

\begin{abstract}
In a previous paper [M.\ E.\ Gusakov, A.\ I.\ Chugunov, and
E.\ M.\ Kantor, Phys.\ Rev.\ Lett.\ {112}, 151101 (2014)],
we introduced a new scenario that explains the existence of
rapidly rotating warm neutron stars (NSs) observed in
low-mass X-ray binaries (LMXBs). Here it is described in
more detail. The scenario takes into account the
interaction between superfluid inertial modes and the
normal (quadrupole) $m=2$ $r$-mode, which can be driven
unstable by Chandrasekhar-Friedman-Schutz (CFS) mechanism.
This interaction can only occur at some fixed ``resonance''
stellar temperatures; it leads to formation of the
``stability peaks'' which stabilize a star in the vicinity
of these temperatures. We demonstrate that a NS in LMXB
spends a substantial fraction of time on the stability
peak, that is, in the region of stellar temperatures and
spin frequencies, that has been previously thought to be
CFS unstable with respect to excitation of $r$-modes. We
also find that the spin frequencies of NSs are limited by
the CFS instability of normal (octupole) $m=3$ $r$-mode
rather than by $m=2$ $r$-mode. This result agrees with the
predicted value of the cutoff spin frequency $\sim 730$~Hz
in the spin distribution of accreting millisecond X-ray
pulsars. In addition, we analyze evolution of a NS after
the end of the accretion phase and demonstrate that
millisecond pulsars can be born in LMXBs within our
scenario. Besides millisecond pulsars, our scenario also
predicts a new class of LMXB descendants---hot and rapidly
rotating nonaccreting NSs (``hot widows''/HOFNARs). Further
comparison of the proposed theory with observations of
rotating NSs can impose new important constraints on the
properties of superdense matter.
\end{abstract}

\pacs{97.60.Jd , 97.80.Jp, 97.60.Gb,95.30.Sf, 26.60.Dd}

\maketitle

\section{Introduction}
\label{Sec:Intro}

Neutron stars (NSs) are the compact  {\it rotating}
objects with a mass $M \sim M_{\odot}$ and radius $R\sim 10$~km
(e.g., Ref.\ \cite{hpy07}).%
\footnote{ The most rapidly rotating NS observed so far is
the millisecond pulsar PSR~J1748-2446ad with the spin
frequency $\nu=716$~Hz \cite{hrsfkc06}.}
Rotation leads to the appearance of the so-called {\it
inertial} oscillation modes in NSs, whose restoring force
is the Coriolis force \cite{unno_et_al_89}. A particular,
but the most interesting class of inertial modes is {\it
$r$-modes} for which (unlike the other inertial modes) the
dominant oscillations are of toroidal type \cite{ak01}. The
remarkable property of $r$-modes is that, neglecting
dissipation, they are subject to a gravitationally driven
Chandrasekhar-Friedman-Schutz (CFS) instability at {\it
arbitrary} spin frequency $\nu$ of a NS
\cite{andersson98,fm98}. An account for dissipative effects
stabilizes the NS to some extent resulting in the
appearance of the ``stability region'' in the
$\nu-T^\infty$ plane, where $T^\infty$ is the redshifted
internal stellar temperature. A typical stability region is
shaded in grey in Fig.\ \ref{Fig_typical_obs} (see Sec.\
\ref{Subsec_obs_stab}); $r$-modes cannot be spontaneously
excited inside this region.

In some cases observations of rapidly rotating NSs in
low-mass X-ray binaries (LMXBs) allow one to measure $\nu$
(e.g., Refs.\ \cite{patruno10,pw12}) and estimate
$T^\infty$ (e.g., Refs.\ \cite{hah11,hdh12,ms13} and Table
\ref{Tab_LMXB_Observ}). It turns out that many of the
rapidly rotating warm sources fall well outside the
stability region, if it is plotted under realistic
assumptions about the properties of superdense matter
\cite{hah11,hdh12}. In fact, calculations show that NSs in
LMXBs can indeed leave the stability region for a while,
but the probability to observe them there is negligibly
small in most cases (see, e.g., Refs.\
\cite{levin99,heyl02} and Sec.\ \ref{Sec_obs_stab}). Thus,
we face a paradox which is usually being explained
following one of the two approaches.

In the first approach one tries,
making some (rather artificial) assumptions,
to enhance damping of $r$-mode oscillations
due to various dissipative mechanisms.
The aim is
to enlarge
the stability region
so that it would contain all the observed sources
(see, e.g., \cite{ak01,hdh12}).

The second approach assumes that some fraction of NSs lies
{\it outside} the stability region, but their spin
frequency $\nu$ and temperature $T^\infty$ are
determined
by two conditions that should be satisfied simultaneously:
($i$) $r$-mode oscillations in these NSs should reach
saturation because of nonlinear interaction with other
inertial modes (see, e.g., \cite{hah11,ms13} and Sec.\
\ref{Subsec_obs_stab}) and ($ii$) all the heat released due
to dissipation of the ``saturated'' $r$-modes should be
radiated away by the neutrino emission. Unfortunately,
these conditions lead to unrealistically small values of
the saturation amplitude $\alpha_{\rm sat} \sim
10^{-9}$--$10^{-6}$, specific to each source
\cite{hah11,ms13}. Such small $\alpha_{\rm sat}$ seem to
contradict the results of Refs.\ \cite{btw07,btw09} (see
also footnote \ref{BW13} below).

Thus, one can conclude that the existence of rapidly
rotating warm NSs remains an open problem
\cite{Andersson_et_al_13}.
A possible solution to this problem was suggested in our
recent paper \cite{gck14_short} and is discussed in more
detail here. Our key idea consists in that to study
evolution of NSs in LMXBs one has to correctly take into
account the resonance interaction between the normal
oscillation $m=2$ $r$-mode and superfluid inertial modes,
which occurs at some fixed values of $T^\infty$ (see Sec.\
\ref{Sec_sflnorm}). Such resonance interaction has been
completely ignored in the literature so far. However, as we
will argue below, it should take place and can dramatically
affect the evolution of rapidly rotating NSs.

First of all, this interaction modifies the stability
region (see Sec.\ \ref{Sec_wind_regime}) and allows us to
suggest an evolution scenario (Sec.\ \ref{Sec_climb}), that
explains all the sources in LMXBs within the standard,
minimal assumptions about the composition and properties of
superdense matter. Moreover, as directly follows from our
scenario, the NS spin frequencies $\nu$ appear to be
bounded by the onset of the octupole $m=3$ oscillation
$r$-mode instability, which corresponds to $\nu \sim
600$--$700$~Hz at $T^\infty\sim 10^8$~K (see Fig.\
\ref{Fig_scenario}). The existence of an upper bound for
$\nu$ can explain the sharp cutoff of the distribution
function for accreting X-ray pulsars at a frequency $\nu
\gtrsim730$~Hz \cite{chakrabarty_etal_03, chakrabarty08}.
If correct, this result presents a strong argument in favor
of the idea of Refs.\ \cite{bildsten98, aks99} that the NS
spin frequency $\nu$ is limited by the $r$-mode
instability. Note, however, that in our scenario $\nu$ is
limited by the octupole $m=3$ $r$-mode rather than by
quadrupole $m=2$ $r$-mode, as it is supposed in Refs.\
\cite{bildsten98, aks99}.

The paper is organized as follows. In Sec.\ \ref{Sec_input}
we discuss the adopted NS model and write out general
equations governing the thermorotational evolution of a NS
in LMXB with allowance for the excitation of normal
$r$-modes. In Sec.\ \ref{Sec_obs_stab} we present the
summary of observations of quiescent temperatures and spin
frequencies for NSs in LMXBs, and demonstrate the problem
with their explanation within the scenarios available in
the literature. In Sec.\ \ref{Sec_sflnorm} we describe and
justify our model of resonance interaction between the
normal and superfluid oscillation modes. In Sec.\
\ref{Sec_wind_regime} we determine the stability region
taking into account the resonance interaction of the normal
$m=2$ $r$-mode and one of the superfluid inertial modes; we
also generalize the equations describing the NS dynamics to
the case when a few oscillation modes are simultaneously
excited in a star. These results are applied in Sec.\
\ref{Sec_climb} to model the evolution of an accreting NS.
Detailed analysis of the evolution tracks allow us to
formulate an original scenario explaining all the existing
data on the spin frequencies and temperatures of NSs in
LMXBs.
In Sec.\ \ref{MSP} we discuss the NS evolution after the
end of the accretion phase. We argue that our scenario can
explain observations of millisecond pulsars and also
predicts the existence of a new possible class of hot,
nonaccreting, and  rapidly rotating NSs. In Sec.\
\ref{Sec_concl} we present the main conclusions.

\section{Physics input and general equations}
\label{Sec_input}

\begin{figure}
    \begin{center}
        \leavevmode
        \includegraphics[width=3.4in]{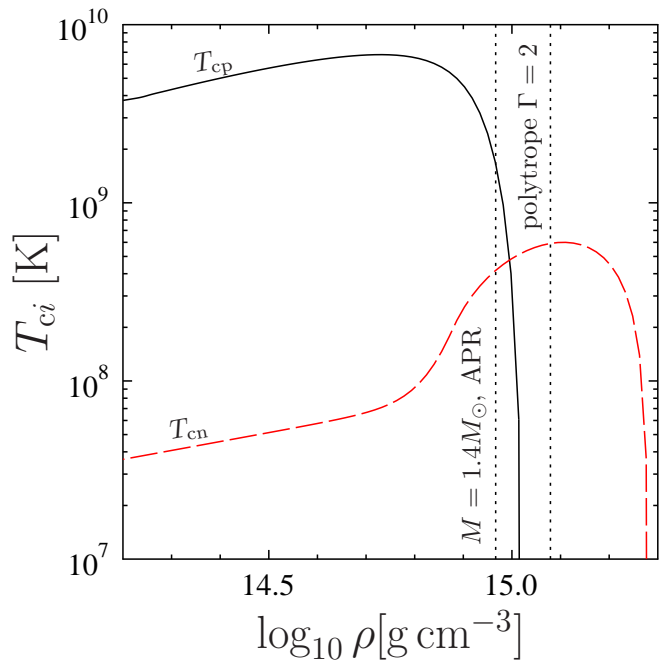}
    \end{center}
    \caption{(color online)
    Critical temperatures of protons $T_{{\rm c}p}$
    (solid line; black online) and neutrons $T_{{\rm c}n}$ (dashed line; red online) as functions of density $\rho$
    in neutron star core.
    Vertical dotted lines indicate
    central densities of a star with the mass $M=1.4M_\odot$.
    The left line corresponds to
    the relativistic star with the Akmal-Pandharipande-Ravenhall (APR) EOS \cite{apr98},
    the right line corresponds to the polytropic Newtonian star
    (polytropic exponent $\Gamma=2$)
    with the radius $R=10$~km.
    }
    \label{Fig_Tci_profile}
\end{figure}

All calculations in this paper are carried out for
a canonical NS with the mass $M=1.4 M_{\odot}$ and radius $R=10$~km,
whose core is composed of neutrons
($n$), protons ($p$) and electrons ($e$).
Following Refs.\ \cite{olcsva98, lom98, hl00, heyl02, wagoner02}
we, for simplicity,
consider the polytropic equation of state (EOS)
with polytropic index $n=1$
($P \propto \rho^\Gamma$,
where $\Gamma=1+1/n =2$; $P$ and $\rho$ are, respectively,
the pressure and density of matter).
We checked,
that use of more realistic EOSs does not affect
our
main results (see also Ref.\ \cite{ak01}).

According to numerous microscopic calculations, nucleons
(neutrons and protons) in the internal layers of NSs are
superfluid at temperatures $T \la 10^8$--$10^{10}$~K.
Recent real-time observations of a cooling young NS in
Cassiopeia A supernova remnant \cite{hh10} have presented
strong evidence of this fact (but see a critique in Ref.\
\cite{Posset_etal13}). They were explained
\cite{shternin11, page11} within the so-called ``minimal
cooling scenario,'' proposed in Refs. \cite{gkyg04,
plps04}. In this paper we use the same models of neutron
and proton superfluidity [that is, the same functions
$T_{{\rm c}i}(\rho)$, where $T_{{\rm c}i}$ is the critical
temperature for transition of a nucleon species $i=n$, $p$
to the superfluid state] as in Ref.\ \cite{gkyg04} (see
Fig.\ \ref{Fig_Tci_profile}); these models are analogous to
those used in Ref.\ \cite{shternin11} to explain the NS
cooling in Cassiopeia A supernova remnant. The
superfluidity models adopted here are also capable of
explaining all observations of cooling isolated NSs
available to date \cite{gkyg04,gkyg05}.

To analyze oscillations of rotating stars it is convenient
to separate the variable $\phi$ that describes the
azimuthal angle in the plane perpendicular to the stellar
rotation axis, and to present all perturbations as $\propto
\exp(\imath m \phi)$, where $m$ is an integer. As it was
shown in Refs.\ \cite{ac01, ly03, yl03a, yl03b, kg13},
inertial modes of two types exist in superfluid NSs for any
$m$.
In Ref.\ \cite{yl03a} they were termed  $i^o$- and $i^s$-modes.%
\footnote{The superscripts $o$ and $s$ here are the
abbreviations for ``ordinary'' and ``superfluid'',
respectively.}
The modes of the first type, which we call ``normal''
($i^o$-modes) describe comoving oscillations of superfluid
and normal matter components and resemble, in many aspects,
the corresponding modes of a normal (nonsuperfluid) star
\cite{gk11, gkcg13, kg13,gkgc14}. The modes of the second
type, which we call ``superfluid'' ($i^{s}$-modes)
correspond to countermoving oscillations of superfluid and
normal matter components and are absent in normal stars. As
it was first demonstrated in Refs.\ \cite{ly03,yl03a}, a
gravitationally driven instability of $i^s$-modes is
strongly suppressed, because their gravitational radiation
is weak, while dissipation of these modes is dramatically
enhanced due to the powerful mutual friction mechanism
(see, e.g., Refs.\ \cite{als84, asc06} and Sec.\
\ref{Subsec_2assum} for more details on the mutual friction
force). Among $i^o$-modes we only consider the normal
$r$-modes with $m=2$ and $m=3$, since they are the most
unstable ones \cite{olcsva98, ak01}. Following Ref.\
\cite{yl03a} we denote the normal $r$-modes as $r^o$-modes;
in this section we analyze them in more detail.

In $r^o$-modes the oscillations are predominantly of toroidal type.
In that case,
to leading order in $\Omega$
(where $\Omega=2 \pi \nu$
is the circular spin frequency),
the Eulerian velocity perturbation
$\delta \pmb{v}$
can be presented as
\cite{pbr81}
\begin{equation}
  \delta {\pmb v} =\alpha \frac{\Omega R r}{\sqrt{l (l+1)}} \left(\frac{r}{R}\right)^l
         {\pmb \nabla} \times(r {\pmb \nabla} Y_{lm}) \mathrm{e}^{\imath \omega t},
         \label{deltaVprofile}
\end{equation}
where $Y_{lm}$ is the spherical harmonic with the
multipolarity $l$ equal to $m$, $l=m$; $\alpha$ is the
oscillation amplitude of the $r^o$-mode; $r$ is the radial
coordinate. Finally, $\omega$ is the oscillation frequency
in the inertial frame, given by (also to leading order in
$\Omega$) \cite{pp78}
\begin{equation}
\omega = - \frac{(l-1) (l+2)}{l+1} \, \Omega.
\label{o}
\end{equation}

Below
we make use of the
quantity
\begin{equation}
\Omega_0 \equiv \sqrt{\pi \, G \, \bar{\rho}}\approx 1.180
\times 10^4\, \left(\frac{M}{1.4M_\odot} \right)^{1/2}
 \left(\frac{R}{10\ \mathrm {km}}\right)^{-3/2}\mathrm s^{-1},
\end{equation}
where $G$ is the gravitation constant and $\bar{\rho}=3M/(4 \pi R^3)$
is the mean stellar density.
For a canonical NS
$\bar{\rho} \approx 6.646 \times 10^{14}$~g~cm$^{-3}$.

To describe the evolution of a NS allowing for the
$r^o$-mode instability, we follow the phenomenological
approach suggested by Owen \textit{et al.} \cite{olcsva98}
and further refined in Refs.\ \cite{hl00} and \cite{as12}.
We mostly employ the notation of Ref.\ \cite{hl00}. The
evolution is given by the following equations:

($i$) An equation governing the variation of canonical
angular momentum $J_{\rm c}$ of the $r^o$-mode due to
radiation of gravitational waves and various dissipative
effects,
\begin{equation}
\frac{d J_{\rm c}}{dt}=
-2 \, J_{\rm c} \,\left(\frac{1}{\tau_{\rm GR}}  + \frac{1}{\tau_{\rm Diss}} \right).
\label{Jc}
\end{equation}
Here \cite{fs78b,olcsva98}
\begin{equation}
   J_{\rm c} =-\frac{l}{2(\omega+l\Omega)}\int \rho\, \delta
             {\pmb v}\, \delta {\pmb v}^\ast d^3 r
             =- \frac{\alpha^2\,l(l+1)}{4}\Omega
            R^{-2l+2}\int_0^R \rho r^{2l+2} d r,
\label{Jc20}
\end{equation}
where
we apply Eqs.\ (\ref{deltaVprofile}) and (\ref{o}) in the second equality.
An integral in the right-hand side of Eq.\ (\ref{Jc20})
can be easily calculated
if one specifies the density profile $\rho(r)$.
Obviously, the integral can generally be written in the form
$\widetilde{J} \, M \, R^{2l}$,
where $\widetilde{J}$
is some numerical coefficient
that depends on $\rho(r/R)$.
Using this expression,
$J_{\rm c}$ can be presented as
\begin{equation}
J_{\rm c}=-\frac{l(l+1)}{4} \, \widetilde{J} \, \, M \, R^2 \, \Omega \, \alpha^2.
\label{Jc2}
\end{equation}
For the simple polytropic model with $\Gamma=2$
and a given stellar mass $M$ and radius $R$, one has
\begin{equation}
  \rho(r)=\frac{M}{4 \, r\,  R^2} \, \sin \left(\frac{\pi r}{R}\right),
  \label{rho_profile}
\end{equation}
which leads to $\widetilde{J} \approx 1.6353\times 10^{-2}$
for $l=m=2$ and $\widetilde{J} \approx 9.9887 \times
10^{-3}$ for the $l=m=3$ $r^o$-mode.

An intensity of gravitational radiation is determined by
the mass current multipole; using Eq.\
(\ref{deltaVprofile}) one can calculate the corresponding
gravitational radiation time scale $\tau_{\rm GR}$
\cite{lom98},
\begin{equation}
\frac{1}{\tau_{\rm GR}} = - \frac{32 \, \pi \, G \, \Omega^{2l+2}}{c^{2l+3}} \,
\frac{(l-1)^{2l}}{[(2l+1)!!]^2} \,
\left( \frac{l+2}{l+1}\right)^{2l+2} \,
\int_0^R \rho \, r^{2l+2} dr,
\label{tauGR}
\end{equation}
where
$c$ is the speed of light.
For the density profile (\ref{rho_profile})
this expression can be rewritten as \cite{ak01}
\begin{equation}
    \tau_{\rm GR} = -\tau_{\rm GR \, 0}
        \left(\frac{M}{1.4 M_\odot}\right)^{-1}
        \,\left(\frac{R}{10\,\mathrm{km}}\right)^{-2l}
        \,\left(\frac{\nu}{1 \mathrm{kHz}}\right)^{-2l-2},
\label{tauGR2}
\end{equation}
where $\tau_{{\rm GR} \, 0}
\approx 46.4$~s and $1250$~s
for $l=m=2$ and $l=m=3$ $r^o$-modes,
respectively.

Further, $1/\tau_{\rm Diss}$ in Eq.\ (\ref{Jc}) is generally presented
in the form,
\begin{equation}
\frac{1}{\tau_{\rm Diss}}=\sum_i \frac{1}{\tau_i},
\label{taudiss}
\end{equation}
where the summation is assumed over all possible processes
resulting in dissipation of energy and angular momentum of
$r^o$-modes (the shear and bulk viscosities, Ekman layer
dissipation, mutual friction etc. \cite{ak01}). In this
paper, we neglect the bulk viscosity, because it is small
for the range of stellar temperatures $T<5 \times 10^8$~K
we are interested in (see, e.g.,
\cite{hly01,gusakov07,kg11,gkcg13}). One can also freely
ignore the effects of mutual friction when considering
$r^o$-modes \cite{lm00,ly03,hap09}. On the opposite,
dissipation in the Ekman layer can be a very efficient
mechanism, though the corresponding damping time $\tau_{\rm
Ek}$ is very sensitive to the chosen model of interaction
between the ``solid'' crust and liquid core of a NS
\cite{lu01, yl01, ak01,
Rieutord01,Rieutord01_erratum,ga06a,km03,mendell01}.
Actually, in the vicinity of the crust-core interface the
crust is neither solid nor liquid, being some intermediate
structure, which is called mantle. Thus, dissipation in the
transition Ekman layer can be substantially lower than it
is often assumed.

Bearing this in mind, we consider dissipation due to the
shear viscosity as our minimal model for the dissipation of
$r^o$-modes. The corresponding time scale $\tau_{\rm S}$
can be calculated from the formula \cite{lom98}
\begin{equation}
\frac{1}{\tau_{\rm S}}=(l-1)(2l+1) \, \int_{0}^R  \eta \, r^{2l} \, dr
\,\, \left( \int_0^R  \rho \, r^{2l+2} dr \right)^{-1},
\label{tauS}
\end{equation}
that was obtained using velocity field
(\ref{deltaVprofile}). Here $\eta$ is the shear viscosity
coefficient. Estimates show that the proton shear viscosity
is small in comparison to the electron one $\eta_e$
\cite{sy08}, while the neutron shear viscosity is poorly
known even for nonsuperfluid NS matter (its value differs
for different authors by a factor of 5--10 and can be
either greater \cite{bv07,zlz10} or smaller
\cite{sy08,sbh13} than $\eta_e$). In view of these facts,
for $\eta$ in this paper we take the electron shear
viscosity $\eta_e$ from Ref.\ \cite{sy08}. Notice that
$\eta_e$ can vary several-fold depending on a chosen EOS
(or, more precisely, depending on a proton fraction
predicted by an EOS; see, e.g., figure 1 in Ref.\
\cite{sy08}). Another important ingredient, affecting
$\eta_e$ \cite{sy08}, is still poorly known model of proton
superfluidity [the profile $T_{{\rm c}p}(\rho)$].

The uncertainties, described above, and possible
contribution of the Ekman layer into dissipation, can {\it
effectively} increase $\eta$ by a factor of few. For
octupole ($l=m=3$) $r^o$-mode the situation is even more
uncertain, because this mode becomes unstable (and thus
important for the NS evolution; see Sec.\
\ref{Sec_wind_regime}) at rather high values of $\Omega$.
This means that the approximation of slowly rotating NSs,
assumed in derivation of Eqs.\ (\ref{tauGR}) and
(\ref{tauS}), can lead to larger errors for the octupole
$r^o$-mode \cite{jas02,kyye00}. Taking this into account,
when modeling the octupole $r^o$-mode (but not the
quadrupole $r^o$-mode!), for $\eta$ we take (somewhat
arbitrary) $\eta_e$ from Ref.\ \cite{sy08}, multiplied by a
factor of $5$; that is, we set $\eta = 5 \eta_e$.

Using the results of Ref.\ \cite{sy08},
we approximate the electron shear viscosity $\eta_e$
by the following
fitting
formula,
\begin{equation}
    \eta_e = 6 \times 10^{18} \,
     \left(\frac{\rho}{10^{15}\ \mathrm g\, \mathrm{cm}^{-3}} \right)^2
     \left(\frac{T}{10^9 \, {\rm K}} \right)^{-2}
     \left(\frac{T_{{\rm c}p}}{2 \times 10^9\, {\rm K}}
     \right)^{1/3} \quad \frac{{\rm g}}{{\rm cm} \, {\rm s}},
     \label{eta}
\end{equation}
which particularly well describes $\eta_e$ for the APR EOS
\cite{apr98} (more precisely, for the parametrization
\cite{hh99} of the APR EOS). Notice that this formula is
valid only if protons are superfluid and $T \la \,0.2
\,T_{{\rm c}p}$. Notice also that, without the last
multiplier, the formula (\ref{eta}) coincides with the
well-known and widely used fit \cite{cl87} of old
calculations of Flowers and Itoh \cite{fi79}. This is an
accidental and surprising coincidence, because the physics
input used in Refs.\ \cite{sy08} and \cite{fi79} is
essentially different (in particular, unlike Ref.\
\cite{sy08}, $\eta_e$ from the paper by Flowers and Itoh
was derived assuming no proton superfluidity and, what is
more important, accounting incorrectly for the effects of
transverse plasma screening on the processes of
electron-electron scattering). In addition, the fitting
formula of Ref.\ \cite{cl87} was obtained for an absolutely
different EOS.

For our model of the proton superfluidity the last
multiplier in Eq.\ (\ref{eta}) is of the order of unity in
the greatest portion of the star, $[T_{{\rm
c}p}(\rho)/(2\times 10^9 \, {\rm K})]^{1/3} \sim 1$. In
view of the uncertainties in the value of $\eta$, we ignore
this multiplier in what follows. Using Eq.\ (\ref{eta}) and
integrating (\ref{tauS}) over $r$, we obtain
\begin{equation}
\tau_{\rm S}=
\tau_{{\rm S} \, 0}
\left(\frac{R}{10\,\mathrm{km}}\right)^{5}
\left(\frac{M}{1.4M_\odot}\right)^{-1} \, \left(T^\infty_8 \right)^2,
\label{tauS2}
\end{equation}
where $T^\infty_8\equiv T^\infty/(10^8\, {\rm K})$;
$\tau_{{\rm S}\, 0} \approx 2.2 \times 10^5$~s for the
$l=m=2$ $r^o$-mode and $\tau_{{\rm S}\, 0} \approx 2.4
\times 10^4$~s for the $l=m=3$ $r^o$-mode (we remind the
reader that in the latter case we take $\eta=5 \eta_e$). In
Eq.\ (\ref{tauS2}), instead of $T$, we introduced the
redshifted internal temperature $T^{\infty} \equiv T \,
{\rm e}^{\nu(r)/2}$, where $\nu(r)$ is the corresponding
metric coefficient \cite{chandrasekhar64}. Let us remind
the reader that in the nonrelativistic approximation which
has been used in derivation of this equation,
$T=T^{\infty}$, so that such replacement is justified.
Moreover, the temperature $T^\infty$, which is constant
over the star, is a more appropriate parameter than $T$ for
the description of NS thermal evolution [see Eq.\
(\ref{thermal}) below] and, especially, for the analysis of
observational data (Sec.\ \ref{Subsec_obs}).

($ii$) An equation describing the change in the total
angular momentum $J_{\rm c}+I \Omega$ of a NS,
\begin{equation}
  \frac{d(J_{\rm c}+I \Omega)}{dt}=-\frac{2}{\tau_{\rm GR}}
   \, J_{\rm c} + \dot J_\mathrm{acc},
\label{Jfull}
\end{equation}
due to gravitational wave radiation (the first term) and
accretion from the low-mass companion (the second term
$\dot J_\mathrm{acc}$). For simplicity, we ignore possible
magnetodipole torque in this paper (but see Sec.\
\ref{Sec_climb}). In Eq.\ (\ref{Jfull}) $I=\widetilde{I} \,
M R^2$ is the stellar moment of inertia; for a polytropic
EOS ($\Gamma=2$) $\widetilde{I} \approx 0.261$. There is a
number of accretion models, leading to somewhat different
estimates for $\dot J_\mathrm{acc}$ (e.g.,
\cite{gl79b,rfs04,kr07}); however, they do not agree well
with observations (see, e.g., \cite{pw12,ajh14}). Thus, for
definiteness, we make use of the simplest estimate,
\begin{equation}
  \dot J_\mathrm{acc}=p \, \dot{M}\, \sqrt{G  M  R},
  \label{Jacc}
\end{equation}
which is traditionally applied in modeling the NS evolution
in binary systems. Here $\dot{M}$ is the mass of accreted
matter per unit time;
$p$ depends on the physics of accretion (i.e., on the NS
magnetic field, spin frequency $\Omega$, accretion rate
etc.; see, e.g., Ref.\ \cite{rfs04}). For simplicity, we
take $p=1$ (e.g., Ref.\ \cite{levin99}).
Below we analyze the large time-scale evolution of NSs;
hence, we assume that the quantities $\dot J_\mathrm{acc}$
and $\dot{M}$ are averaged over the active and quiescent
phases of accretion. Since $\dot J_\mathrm{acc} \propto
\dot{M}$ in Eq.\ (\ref{Jacc}), one can use that expression
for the averaged values as well. In what follows we set
$p=1$ and $\dot{M}=3.0 \times 10^{-10} \, M_\odot$
yr$^{-1}$. The chosen value of $\dot{M}$ is close to the
estimates of the accretion rates for the sources SAX
J1750.8-2900 and 4U 1608-522 (see below).

($iii$) An equation describing the thermal evolution of an
oscillating star,
\begin{equation}
C_{\rm tot} \frac{dT^\infty}{dt} = W_{\rm Diss}-L_{\rm cool} + K_{\rm n} \dot{M} c^2,
\label{thermal}
\end{equation}
where $W_{\rm Diss}$ is the energy
dissipated per unit time
due to the $r^o$-mode damping.
It is presented as (e.g., Ref.\ \cite{levin99})
\begin{equation}
W_{\rm Diss} = \frac{2 E_{\rm c}}{\tau_{\rm Diss}}
= \frac{\widetilde{J}  M R^2  \Omega^2  \alpha^2}{ \tau_{\rm Diss}},
\label{Wdiss}
\end{equation}
where $E_{\rm c}$ is the canonical energy of the $r^o$-mode
(with arbitrary $m$) in a reference frame, rotating with
the star. As it was shown in Refs.\ \cite{fs78a, fs78b},
$E_{\rm c}$ is related to the canonical angular momentum
$J_c$ [see Eq.\ (\ref{Jc2})] by
\begin{equation}
E_{\rm c} = - \frac{(\omega + m \Omega)}{m}\, J_{\rm c}.
\label{Ec}
\end{equation}
This relation is valid for any inertial modes (not only for
$r^o$-modes). Further, $C_{\rm tot}(T^{\infty})$ in Eq.\
(\ref{thermal}) is the total heat capacity of a NS; $L_{\rm
cool}(T^{\infty})$ is its luminosity, that is, the energy
carried away from the star per unit time in the form of
neutrino and electromagnetic radiation from its surface.
Since oscillation amplitudes of $r^o$-modes, analyzed in
this paper, are small ($\alpha \leq 10^{-4}$, see below),
$L_{\rm cool}$ is given by the same equation as for a
nonoscillating star \cite{gyg05}. To determine the
quantities $C_{\rm tot}$ and $L_{\rm cool}$ as accurately
as we can, we calculate them with the relativistic cooling
code, described in detail in Refs.\
\cite{gkyg04,gkyg05,yp04} (we used essentially the same
microphysics input as that employed in Ref.\
\cite{gkyg04}). In particular, we used the parametrization
\cite{hh99} of the APR EOS \cite{apr98} and considered a
star with the mass $M=1.4 M_{\odot}$. Although this
approach is somewhat inconsistent (other equations neglect
relativistic effects and employ the polytropic EOS), it
allows us to use the realistic values for $C_{\rm tot}$ and
$L_{\rm cool}$ in our simplified model. The calculations of
$C_{\rm tot}$ and $L_{\rm cool}$ have been roughly
approximated as functions of internal (redshifted) stellar
temperature $T^\infty$ and are presented in Appendix
\ref{Appendix_LC}. Since the photon luminosity is not
important in the temperature range of interest to us
($T^\infty >10^8$~K), we fit only the neutrino luminosity
in Appendix \ref{Appendix_LC}. Note that for lower
$T^\infty$ the photon luminosity rapidly becomes the main
cooling agent and hence cannot be ignored \cite{yp04}. We
have checked, that the results for $r^o$-mode evolution
obtained using the fitting formulas from Appendix
\ref{Appendix_LC}, practically do not differ from those
obtained using the exact values for $C_{\rm tot}$ and
$L_{\rm cool}$.

Finally, the last term in Eq.\ (\ref{thermal}) describes
the stellar heating due to accretion (deep crustal heating,
see, e.g., Ref.\ \cite{bbr98}). Under the pressure of
accreted material, the matter in the stellar envelope
compresses and eventually undergoes a set of exothermal
nuclear transformations (pycnonuclear reactions and
reactions of beta-capture, accompanied by the neutron
emission). The heat released in these reactions is mostly
accumulated by the core due to high thermal conductivity of
the internal layers of NSs. The parameter $K_{\rm n}$
characterizes the efficiency of this heating; following
Refs.\ \cite{brown00, btw07} we adopt $K_{\rm n}=10^{-3}$
as a fiducial value.
\footnote{$K_{\rm n}=10^{-3}$ corresponds to
the total deep crustal heat release
$\sim1$~MeV per accreted nucleon. Recent calculations
\cite{hz08} suggest a larger value ($\sim 1.5-1.9$~MeV per
accreted nucleon), and even this heat release seems to be
insufficient for explaining crust thermal relaxation of
some LMXBs after an accretion episode (see, e.g., Refs.\
\cite{Shternin_additional_heating,degenaar_etal14}).
However, the actual value of $K_{\rm n}$ is rather
unimportant for our scenario and cannot change our results
qualitatively. }
For a chosen NS model, the heating (in the absence of a
$r^o$-mode) is completely compensated by the cooling
($L_{\rm cool}=K_{\rm n} \dot{M} c^2$) at $T_{\rm
eq}^\infty \approx 1.078 \times 10^8$~K.

Equations (\ref{Jc}), (\ref{Jfull}), and (\ref{thermal})
fully describe the evolution of nonsaturated $r^o$-modes.
Using Eqs.\ (\ref{Jc}) and (\ref{Jfull}) one can express
the quantities $d \alpha/dt$ and $d \Omega/dt$,
\begin{eqnarray}
\frac{d \alpha}{dt} &=& -\alpha \left( \frac{1}{\tau_{\rm GR}}+\frac{1}{\tau_{\rm Diss}} \right),
\label{dadt}\\
\frac{d \Omega}{dt} &=& - \frac{2 \, Q \, \alpha^2 \, \Omega}{\tau_{\rm Diss}}+\dot{\Omega}_{\rm acc},
\label{dOmegadt}
\end{eqnarray}
where
\begin{eqnarray}
\dot{\Omega}_{\rm acc} &\equiv& \dot J_\mathrm{acc}/I=p \,
\dot{M}\, \frac{\sqrt{G  M  R}}{I}
\nonumber\\
&\approx&
3.73 \times 10^{-6} \,\, p \,\, \dot{M}_{-10} \,\, \widetilde{I}_{0.261}^{-1} \,\,
\left(\frac{M}{1.4 M_{\odot}}\right)^{-1/2} \,
\left(\frac{R}{10 \, {\rm km}}\right)^{-3/2} \, {\rm s^{-1} \, yr^{-1}},
\label{Omega_acc}\\
Q &\equiv&
\frac{l(l+1) \widetilde{J} }{4 \widetilde{I}}, \label{Q}
\end{eqnarray}
and $\dot{M}_{-10}=\dot{M}/(10^{-10} \, M_{\odot}\, {\rm
yr}^{-1})$, $\widetilde{I}_{0.261}=\widetilde{I}/0.261$. In
deriving Eq.\ (\ref{dadt}) we neglected the term $\propto
\alpha^3$, assuming that $\alpha \ll 1$. In addition,
because $\dot{\Omega}_{\rm acc}/\Omega \ll 1/\tau_{\rm GR}$
we also neglected the term proportional to
$\dot{\Omega}_{\rm acc}/\Omega$ in Eq.\ (\ref{dadt}). Let
us note that the explicit dependence of the accretion
torque on an accretion regime and its parameters ($\dot M$,
magnetic field etc.) is not important for the final
equations, because they only depend on the accretion torque
$\dot{\Omega}_{\rm acc}$, averaged over a large period of
time, containing both the active and quiescent phases.
In principle, $\dot{\Omega}_{\rm acc}$ can include also
additional braking/spin-up mechanisms which are not related
to the $r$-modes (magnetodipole braking, for example).

The resulting Eqs.\ (\ref{thermal}), (\ref{dadt}), and
(\ref{dOmegadt}) correctly describe the NS evolution only
until a growing oscillation mode enters the nonlinear
saturation regime, where it will interact nonlinearly with
other inertial modes. Under some simplifying assumptions
the nonlinear regime was studied in Refs.\
\cite{schenk_et_al_01, arras_et_al_03, btw04a, btw04b,
btw05, btw07, btw09}. In particular, in the recent papers
by Bondarescu \textit{et al.}\ \cite{btw07,btw09} it has
been shown that the saturation amplitude $\alpha_{\rm sat}$
for the $r^o$-mode can be rather small,
 $\alpha_{\rm sat} \sim
10^{-4}$--$10^{-1}$. Unless otherwise stated, we, following
Ref.\ \cite{btw07}, assume that $\alpha_{\rm sat}=10^{-4}$
for all modes considered in this paper.%
\footnote{We note that the $r^o$-mode amplitude $C_{R}$ of
Bondarescu \textit{et al.} is related to our amplitude
$\alpha$ by $C_{R} = (\widetilde{J}/2)^{1/2} \, \alpha
\approx 0.1 \alpha$, see the footnote 1 in Ref.\
\cite{btw09}. \label{bondarescu_ampl}}

We also assume, as in Ref.\ \cite{olcsva98}, that in the
saturation regime (when $\alpha$ reaches the value
$\alpha_{\rm sat} =10^{-4}$) the oscillation amplitude
stops to grow, so that the energy, pumped into the
$r^o$-mode by gravitational radiation, redistributes among
the other modes through the nonlinear interactions, and
eventually dissipates into heat. Mathematically this can be
(qualitatively) described by introducing in Eq.\
(\ref{dadt}) the effective dissipation time $\tau_{\rm
Diss}^{\rm eff}$ instead of $\tau_{\rm Diss}$, and
requiring that $d \alpha/dt=0$,
\begin{equation}
\frac{d \alpha}{dt}=0=- \alpha \left(\frac{1}{\tau_{\rm GR}}+\frac{1}{\tau_{\rm Diss}^{\rm eff}} \right),
\label{dadt_sat}
\end{equation}
which leads to
\begin{equation}
\tau_{\rm Diss}^{\rm eff}=-\tau_{\rm GR}.
\label{tau2}
\end{equation}
In conclusion, in the saturation regime we ($i$) fix the
amplitude of the $r^o$-mode $\alpha=\alpha_{\rm
sat}=10^{-4}$, and ($ii$) replace $\tau_{\rm Diss}$ with
$\tau_{\rm Diss}^{\rm eff}=-\tau_{\rm GR}$ in Eqs.\
(\ref{thermal}) and (\ref{dOmegadt}). Let us notice that,
when modeling the saturated oscillations, Owen \textit{et
al.} \cite{olcsva98} did not replace the quantity
$\tau_{\rm Diss}$ in the thermal evolution equation
(\ref{thermal}) [but replaced it in Eq.\ (\ref{dOmegadt})].
The authors of Ref.\ \cite{as12} were the first to
emphasize that it would be more self-consistent to replace
$\tau_{\rm Diss}$ with $-\tau_{\rm GR}$ also in Eq.\
(\ref{thermal}).

\section{Observational data and stability of rapidly rotating neutron stars}
\label{Sec_obs_stab}

\subsection{Observational data}
\label{Subsec_obs}

\begin{table}[h!]
\caption{Observational data and internal temperatures on
NSs in LMXBs}
\begin{tabular}{l r r r r r r r r}
\hline
   Source               &  $\nu$ $[\mathrm{Hz}]$
   & $\displaystyle \frac{T^\infty_{\mathrm{eff}}}{10^6\, \mathrm K}$
   & Ref.\
   & $\displaystyle \frac{T^\infty_{\mathrm{acc}}}{10^8\, \mathrm K}$
   & $\displaystyle \frac{T^\infty_{\mathrm{fid}}}{10^8\, \mathrm K}$
   & $\displaystyle \frac{T^\infty_{\mathrm{Fe }}}{10^8\, \mathrm K}$
   & $\displaystyle \frac{\dot M}{M_\odot}$ $[\mathrm{yr^{-1}}]$
   & Ref.\ \\
   \hline\hline
   4U 1608-522          &    $620$    &   $1.51\,\ $        & \cite{rutledge_et_al_99} &   $0.93$        & $1.90$        &    $2.47$        & $3.6\times10^{-10}$& \cite{hjwt07}\\
   SAX J1750.8-2900     &    $601$    &   $1.72\,\ $        & \cite{lowell_et_al_12}   &   $1.18$        & $2.57$        &    $3.11$        & $2\times10^{-10}$& \cite{lowell_et_al_12}\\
   IGR J00291-5934      &    $599$    &   $0.63$\footnote{We treat the effective temperature from Table 2
    of Ref.\ \cite{heinke_et_al_09} as a local one to reproduce the thermal luminosity from that reference.
\label{Heinke_Ts_correction}}
                                                        & \cite{heinke_et_al_09}   &   $0.21$        & $0.24$        &    $0.52$        & $2.5\times10^{-12}$& \cite{heinke_et_al_09}\\
   MXB 1659-298         &    $567$\footnote{According to Refs.\ \cite{wsf01,watts_et_al_08,watts12}}
                                      &   $0.63\,\ $       & \cite{cackett_et_al_08}  &   $0.21$        & $0.24$        &    $0.52$        & $1.7\times10^{-10}$& \cite{hjwt07}\\
   EXO 0748-676 \footnote{The radius of this source was fixed at 15.6~km in spectral fits of Ref.\ \cite{degenaar_et_al_11}.}
                        &    $552$    &   $1.26\,\  $       & \cite{degenaar_et_al_11} &   $0.68$        & $1.20$        &    $1.79$        & \\
   Aql X-1              &    $550$    &   $1.26\,\  $       & \cite{cackett_et_al_11}  &   $0.68$        & $1.20$        &    $1.79$        & $4\times10^{-10}$& \cite{hjwt07}\\
   KS 1731-260          &    $524$\footnote{According to Refs.\ \cite{muno_et_al_00,watts_et_al_08,watts12}}
                                      &   $0.73\,\  $       & \cite{cackett_et_al_10}  &   $0.27$        & $0.32$        &    $0.67$        & $<1.5\times10^{-9}$& \cite{hjwt07}\\
   SWIFT J1749.4-2807   &    $518$    &   $<1.16\,\ $       & \cite{dpw12}             &   $0.59$        & $0.96$        &    $1.54$        & \\
   SAX J1748.9-2021     &    $442$    &   $1.04\,\ $       & \cite{cackett_et_al_05}  &   $0.49$        & $0.72$        &    $1.27$         & $1.8\times10^{-10}$& \cite{hjwt07}\\
   XTE J1751-305        &    $435$    &   $<0.63$\footnotemark[1]     & \cite{heinke_et_al_09}
                                                                                   &   $0.21$        & $0.24$        &    $0.52$        & $6\times10^{-12} $& \cite{heinke_et_al_09}\\
   SAX J1808.4-3658     &    $401$    &   $<0.27$\footnotemark[1]& \cite{heinke_et_al_09}
                                                                                   &   $0.05$        & $0.05$        &    $0.11$        & $9\times10^{-12}$& \cite{heinke_et_al_09}\\
   IGR J17498-2921      &    $401$    &   $<0.93\,\ $       & \cite{dpw12}            &   $0.41$        & $0.55$        &    $1.04$        & \\
   HETE J1900.1-2455    &    $377$    &   $<0.65\,\ $       & \cite{hdh12}             &   $0.22$        & $0.25$        &    $0.55$       & \\
   XTE J1814-338        &    $314$    &   $<0.61$\footnotemark[1] & \cite{heinke_et_al_09}                                                                                    &   $0.20$        & $0.22$        &    $0.49$         & $3\times10^{-12}$& \cite{heinke_et_al_09}\\
   IGR J17191-2821      &    $294$    &   $<0.86\,\ $       & \cite{hdh12}             &   $0.36$        & $0.45$        &    $0.90$        \\
   IGR J17511-3057      &    $245$    &   $<1.1 \,\ $      & \cite{hdh12}              &   $0.54$        & $0.84$        &    $1.40$        \\
   NGC 6440  X-2        &    $205$
                                      &   $<0.37\,\ $       & \cite{hdh12}             &   $0.09$        & $0.09$        &    $0.20$        & $1.3\times10^{-12}$& \cite{heinke_et_al_10}\\
   XTE J1807-294        &    $190$    &   $<0.45$\footnotemark[1]
                                                        & \cite{heinke_et_al_09}   &   $0.12$        & $0.13$        &    $0.28$       & $<8\times10^{-12}$& \cite{heinke_et_al_09}\\
   XTE J0929-314        &    $185$    &   $<0.58\,\ $       & \cite{wijnands_et_al_05} &   $0.19$        & $0.20$        &    $0.45$        & $<2\times10^{-11}$& \cite{heinke_et_al_09}\\
   Swift J1756-2508     &    $182$    &   $<0.96\,\ $       & \cite{hdh12}             &   $0.43$        & $0.59$        &    $1.10$       &\\
             \hline
\end{tabular}

 \label{Tab_LMXB_Observ}
\end{table}

Observational data on spin frequencies, quiescent
temperatures, and accretion rates are summarized in Table
\ref{Tab_LMXB_Observ} for 20 neutron stars in LMXBs. The
source names are given in the first column. The second
column presents the NS spin frequencies $\nu$ which are
mainly taken from Ref.\ \cite{patruno10}. An exception is
the source IGR J17498-2921, for which we adopt the value of
$\nu$ from the review \cite{pw12}. The third column
summarizes observational data on NS redshifted effective
temperatures $T_{{\rm eff}}^\infty$ in the quiescent state.
The corresponding values are taken from the papers quoted
in the fourth column. In those papers the thermal component
was fitted by the hydrogen atmosphere models with the
fiducial value of the NS mass $M=1.4 M_{\odot}$. Except for
the sources EXO 0748-676 and 4U 1608-522, the NS
circumferential radii were also fixed at the fiducial value
$R=10$~km. In Ref.\ \cite{rutledge_et_al_99} the apparent
emission area radius $r_e$ for the source 4U 1608-522 was
treated as a free parameter, and the value
$r_e=9.4_{-2.7}^{+4.3}$~km, extracted from the spectral
fitting, is compatible with the fiducial value $R=10$~km.
At the same time, the spectral fitting for EXO 0748-676
with the canonical mass $M=1.4 M_{\odot}$ and radius
$R=10$~km leads to unrealistic estimates of the distance
and/or hydrogen column density $N_\mathrm{H}$
\cite{degenaar_et_al_11}, which made the authors of that
reference to fix the radius at the best-fit value
$R=15.6$~km.
\footnote{Slightly different X-ray spectral fits have been
suggested in a recent paper \cite{degenaar_etal14}.
However, the difference
in the fitting parameters is negligible in comparison with
uncertainties related to unconstrained crust composition.}
Let us also note that we treat the values of the effective
temperatures shown in Table\ 2 of Ref.\
\cite{heinke_et_al_09} as the {\it local} (nonredshifted)
ones to reproduce the objects'' thermal luminosities,
calculated in the same paper.%
\footnote{For the source XTE J1751-305 we reproduce an
upper limit of $2\times 10^{32}$~erg\,s$^{-1}$ for the
thermal luminosity obtained in Ref.\
\cite{wijnands_et_al_05}, rather than the value $4\times
10^{32}$~erg\,s$^{-1}$ shown in the Table 2 of Ref.\
\cite{heinke_et_al_09}.}
It is interesting, that the
parameters of the sources EXO 0748-676 and Aql X-1 almost
coincide in Table \ref{Tab_LMXB_Observ}

For each $T_{{\rm eff}}^\infty$ we calculate the internal
redshifted temperature $T^\infty$ by employing the
analytical fitting formulas from Ref.\ \cite{pcy97} (see
Appendix A3 of that reference), and assuming canonical
values of mass and radius for each source (including
EXO~0748-676). The relation between $T_{{\rm eff}}^\infty$
and $T^\infty$ depends on the amount of material accreted
onto the NS surface. To get an impression about uncertainty
in the value of $T^\infty$ at a fixed effective temperature
$T_{\rm eff}^\infty$ we, following Ref.\ \cite{hdh12},
consider three models of envelope composition, ($i$) fully
accreted envelope (the corresponding internal temperature
$T^{\infty}_{\rm acc}$ is given in the fifth column of
Table \ref{Tab_LMXB_Observ}); ($ii$) partially accreted
envelope with a layer of accreted light elements down to a
column depth of $P/g=10^9$~g~cm$^{-2}$ (the corresponding
``fiducial'' temperature $T^\infty_{\rm fid}$ is presented
in the sixth column; $P$ is the pressure at the bottom of
the accreted column, $g$ is the gravitational acceleration
at the stellar surface; the same fiducial value of $P/g$
has been considered in Refs.\ \cite{bc09, hdh12}); ($iii$)
pure iron envelope (the corresponding temperature
$T^\infty_{\rm Fe}$ is given in the seventh column). For
all sources $T^\infty_{\rm acc}<T^\infty_{\rm
fid}<T^{\infty}_{\rm Fe}$, because the thermal conductivity
of the pure iron envelope is lower than that of the
envelope with an admixture of light elements (the iron
envelope is better heat insulator). Note, however, that
this inequality (and its explanation) is only justified at
not-too-low temperatures $T_{\rm eff} \ga 10^5$~K
\cite{pcy97}.

Finally, the eighth column presents estimates of the
{\it averaged} accretion rates $\dot{M}$ onto NSs
and the corresponding references.
The averaging is performed over a long period of time,
which includes both active
and quiescent phases.
Unfortunately, we have not found
estimates of $\dot{M}$ for some sources.

\subsection{Observational data vs stability of rapidly rotating NSs}
\label{Subsec_obs_stab}

%
\begin{figure}
    \begin{center}
        \leavevmode
        \includegraphics[width=3.4in]{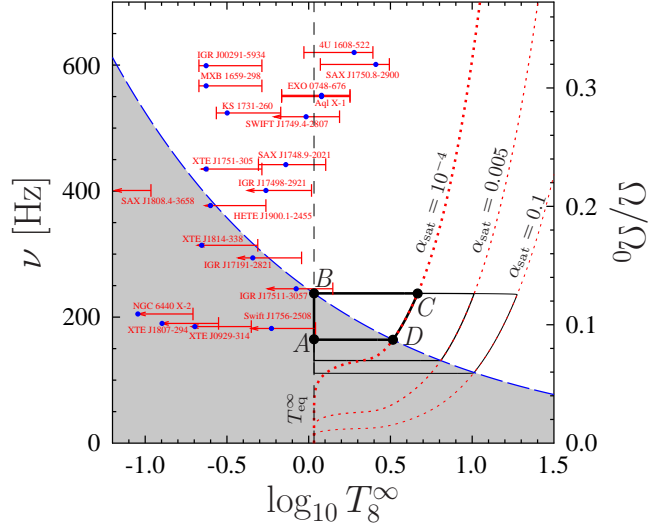}
    \end{center}
    \caption{(color online)
    Spin frequency vs internal redshifted temperature for NSs in LMXBs.
    The frequencies and fiducial temperatures of 20 sources
    from Tab.\ \ref{Tab_LMXB_Observ}
    are shown by small filled circles.
    Error bars
    describe
    uncertainties
    in $T^\infty$
    related to poorly constrained envelope composition
    (see Sec.\ \ref{Subsec_obs} and Table \ref{Tab_LMXB_Observ}).
    Evolution tracks
    for a NS in LMXB are plotted by the solid  lines
    (black online; thick, medium, and thin lines are for
    $\alpha_\mathrm{sat}=10^{-4}$, $0.005$, and $0.1$, respectively).
    Four points $A$, $B$, $C$, and  $D$
    separate different stages of NS evolution on the track,
    which corresponds to $\alpha_\mathrm{sat}=10^{-4}$.
    The stability region for $r^o$-mode with
    $m=2$ is filled with grey,
    its boundary is shown by thick dashed line (blue online).
    The vertical dashed line demonstrates
    the equilibrium stellar temperature $T_\mathrm{eq}^\infty$.
    The dotted lines (red online) are the Cooling=Heating curves
    for $\alpha_\mathrm{sat}=10^{-4}$, $0.005$, and $0.1$.
    See text for details.
    }
    \label{Fig_typical_obs}
\end{figure}
%

The region of typical temperatures and spin frequencies for
NSs in LMXBs is shown in Fig.\ \ref{Fig_typical_obs}. The
small filled circles demonstrate the fiducial temperatures
$T^\infty_{\rm fid}$ of the sources from Table
\ref{Tab_LMXB_Observ}, corresponding to the column depth of
light elements $P/g=10^9$~g~cm$^{-2}$. The error bars
indicate uncertainties in the internal temperature, which
can vary from $T_{\rm acc}^\infty$ (fully accreted
envelope) to $T_{\rm Fe}^\infty$ (iron envelope), see Table
\ref{Tab_LMXB_Observ}. If only an upper limit for the
effective temperature is known for a source, then the left
error bar ends with arrow and the values of $T^\infty_{\rm
fid}$, $T^\infty_{\rm acc}$, and $T^\infty_{\rm Fe}$ are
calculated for that upper limit. Note that, because $\nu$
and $T^\infty_{\rm eff}$ for the sources EXO 0748-676 and
Aql X-1 are very close to one another, the corresponding
error bars almost merge in Fig.\ \ref{Fig_typical_obs}.

By dashes we plot the ``instability curve'' for the
quadrupole $m=2$ $r^o$-mode, which is determined by the
condition $1/\tau_{\rm GR}+1/\tau_{\rm Diss}=0$. Above this
curve $1/\tau_{\rm GR}+1/\tau_{\rm Diss}<0$ and, as follows
from Eq.\ (\ref{dadt}), a star becomes unstable with
respect to excitation of the $r^o$-mode ($d \alpha/dt>0$).
This region is often referred to as the {\it instability
window} for $r$-modes \cite{ak01}. The region filled with
grey in the figure is the stability region for the $m=2$
$r^o$-mode. One can observe that a number of NSs appears
well beyond the stability region.

As it was first shown by Levin \cite{levin99} (see also
Ref.\  \cite{heyl02}), NSs in LMXBs can undergo a cyclic
evolution. This results in a closed track in the
$\nu-T^\infty$ plane with a part of the track belonging to
the instability region. For the NS model described in Sec.\
\ref{Sec_input} and the $r^0$-mode saturation amplitude
$\alpha_{\rm sat}=10^{-4}$ such a track
$A$--$B$--$C$--$D$--$A$ is shown in Fig.\
\ref{Fig_typical_obs} by the thick solid line (black
online); medium and thin solid lines demonstrate similar
tracks for $\alpha_{\rm sat}=5 \times 10^{-3}$ and
$\alpha_{\rm sat}=10^{-1}$, respectively. It is worth
noting that, qualitatively, the shape of these tracks does
not depend on the details of microphysics input adopted in
Sec.\ \ref{Sec_input}.

The evolution tracks in Fig.\ \ref{Fig_typical_obs} consist
of four main stages. Let us describe them briefly, taking
the $A$--$B$--$C$--$D$--$A$ track as an example (a detailed
discussion with a number of useful estimates can be found
in Appendix \ref{Appendix_StandScen}):

($i$) Spin-up of the star in the stability region at a
temperature $T_A^\infty=T_{\rm eq}^\infty$ (stage
$A$--$B$).

The star stays in the stability region and $r^o$-modes are
not excited ($\alpha=0$). In accordance with Eq.\
(\ref{dOmegadt}), the spin frequency increases linearly
with time due to accretion of matter onto the NS, while the
stellar temperature $T^\infty$, governed by Eq.\
(\ref{thermal}), stays constant. This stage lasts
$\tau_{AB}\approx 4\times 10^7$~yr and ends by crossing the
instability curve.

($ii$) Runaway heating of the star in the instability
region (stage $B$--$C$).

This stage starts when the star leaves the $m=2$ $r^o$-mode
stability region due to accretion-driven spin-up. The
corresponding oscillation amplitude $\alpha$ begins to
increase rapidly from the initial value determined by
fluctuations (for example, the thermal fluctuations or
those, related with accretion). Even at very low initial
amplitude $\alpha=10^{-30}$ it takes $\Delta t_{\rm torq}
\approx 4500$~yr for the torque associated with viscous
damping of the $r^o$-mode to become equal to the accretion
torque [$d \Omega/dt =0$, see Eq.\ (\ref{dOmegadt})]. In
the next $\approx 4$~yr, the $r^o$-mode reaches saturation
($\alpha=\alpha_{\rm sat}$). During these two periods of
time, $T^\infty$ and $\Omega$ remain almost unchanged (the
shift of the star in Fig.\ \ref{Fig_typical_obs} is smaller
than the width of the evolution track line).

Having reached saturation, the amplitude of $m=2$
$r^o$-mode stops growing and the star (within the time
$\Delta t_T \approx 3000$~yr) warms up to the temperature,
at which the neutrino emission exactly compensates the
heating caused by the dissipation of the saturated
oscillation mode [see Eq.\ (\ref{thermal})],
\begin{equation}
-\frac{\widetilde{J}  M R^2  \Omega^2  \alpha_{\rm
sat}^2}{\tau_{\rm GR}} -L_{\rm cool} + K_{\rm n} \dot{M} c^2=0.
\label{Coolheat}
\end{equation}
The temperatures that satisfy this condition strongly
depend on the stellar spin frequency and the saturation
amplitude. We will refer to the corresponding curves in the
$\nu-T^\infty$ plane as the Cooling=Heating curves; they
are shown in Fig.\  \ref{Fig_typical_obs} for $\alpha_{\rm
sat}=10^{-4}$, $5 \times 10^{-3}$, and $10^{-1}$ by the
dotted lines (red online). These lines constrain the region
of temperatures and frequencies accessible for NSs in
LMXBs; the star cannot intersect the Cooling=Heating curve
during its runaway, since this requires a more intensive
heating than the dissipation of the saturated mode can
provide. Note that the frequency remains almost unchanged
during the $B$--$C$ stage.

($iii$) Spin-down of the star along the Cooling=Heating
curve in the instability region (stage $C$--$D$).

Having reached point $C$, the star starts to move along the
Cooling=Heating curve; that is, its temperature is
determined by the balance of neutrino luminosity and
heating due to dissipation of the saturated mode. As the
rate of the angular momentum loss associated with the
emission of gravitational waves is larger than the
accretion torque in our NS model,
the star starts to spin down.%
\footnote{For lower saturation amplitudes, the latter
condition may be violated. In that case the star moves to
the stationary point at the Cooling=Heating curve, where
the accretion torque is balanced by the angular momentum
loss due to emission of gravitational waves from the
unstable oscillation mode. }
Eventually, the star returns into the stability region.
This stage lasts $\Delta t_{CD} \approx 8 \times 10^6$~yr.

($iv$) Cooling of the star in the stability region (stage
$D$--$A$).

Having entered into the stability region, the $r^o$-mode
amplitude vanishes rapidly (in $\sim 400$~yr), and after
that a cooling of the star down to the temperature $T_{\rm
eq}^\infty$ (point $A$) takes place. The cooling lasts
$\sim 10^5$~yr, then the cycle repeats. The spin frequency
does not change noticeably during the $D$--$A$ stage.

Summarizing, the star spends most of the time in stage
($i$) and only rarely gets into the instability region.
Furthermore, in the instability region the star spends in
stage ($ii$) a few orders of magnitude less time than in
stage ($iii$).

Obviously, none of the observed NSs in LMXBs evolves along
the tracks in Fig.\ \ref{Fig_typical_obs}. Various
modifications of the standard scenario described above, for
example, decreasing of $T_{\rm eq}^\infty$ (with the aim to
increase $\Omega_{B}$) and increasing of the saturation
amplitude $\alpha_{\rm sat}$, can allow one to interpret
the observed sources as moving along the horizontal part of
the evolution track that corresponds to stage
($ii$)---runaway heating of a star in the instability
region. However, such modifications would make the
detection of any source in this stage even more unlikely
since they would further decrease the fraction of time
spent there by the star \cite{heyl02}. In addition, this
interpretation of observations would also suggest that a
significant number of NSs in LMXBs should be located in
stage ($iii$) (on the Cooling=Heating curve), since the
duration of this stage is a few orders of magnitude larger
than that of stage ($ii$) (see Appendix
\ref{Appendix_StandScen}). As follows from Fig.\
\ref{Fig_typical_obs}, the Cooling=Heating curves (the
dotted lines; red online) correspond to very high
temperatures ($T^\infty\sim 4 \times 10^8$~K), so such
stars should have been observed. Nevertheless, {\it none}
of the NSs detected in LMXBs has a redshifted effective
temperature larger than $T_{\mathrm{eff}}^\infty\gtrsim
2\times 10^6$~K (which corresponds to $T^\infty_{\rm
fid}\gtrsim 4\times 10^8$~K
for the canonical NS model).
\footnote{ Note that in reality it is very difficult to
further increase $T_{\mathrm{eff}}^\infty$ by increasing
$T^\infty$. The reason is a very strong neutrino cooling in
the NS crust which prevents $T_{\mathrm{eff}}^\infty$ from
being larger than a few times $10^6$~K even for
$T^\infty\gtrsim10^9$~K (see, e.g., Ref.\ \cite{pcy07}). }

In other words, the NS temperatures and frequencies
inferred from the LMXB observations cannot be explained
within the standard scenario. Therefore, to explain the
sources from Fig.\ \ref{Fig_typical_obs} one usually
follows a different approach, trying to raise the
instability curves so that all the sources would be
contained inside the stability region. To this aim one
needs to enhance dramatically the dissipation of the $m=2$
$r^o$-mode. Unfortunately, it is very difficult to justify
such an enhancement from the microphysics point of view
\cite{hah11, hdh12}.

An alternative approach to the explanation of the sources
with high temperatures and frequencies was suggested in
Refs.\ \cite{hah11,hdh12,ms13, bw13}. It is based on the
assumption that the NSs observed in the instability region
are in the quasistationary state, in which the stellar
temperature keeps constant ($dT^\infty/dt=0$) by balancing
the neutrino cooling and heating associated with the
dissipation of the saturated $r^o$-mode. However, to
satisfy this condition the saturation amplitudes should
differ substantially from source to source and, in
addition, should have very low values of $\alpha_{\rm
sat}\sim 10^{-9}$--$10^{-6}$, in disagreement with the
recent calculations \cite{btw07,btw09}. According to the
model of Refs.\ \cite{btw07,btw09}, the saturation
amplitude is determined by the lowest parametric
instability threshold among various triplets of the
$r^o$-mode and two inertial {\it daughter} modes with which
it is coupled {\it nonlinearly}. The threshold depends on
the detuning of frequencies in the mode triplet and on the
damping time scales of daughter modes. Since the mode
frequencies are nonlinear functions of the spin frequency
$\nu$, for some $\nu$ a very small detuning can
occasionally occur for not very high daughter modes with
relatively weak damping. It may thus lead to a very low
saturation amplitude. However, such situation seems to be
{\it unstable}, because the variation of the spin frequency
increases detuning and the saturation
amplitude and, as a result, additionally heats up the star.
%
\footnote{ In a recent paper \cite{bw13} it is argued that
the {\it minimum} saturation amplitude $|C_R|_{\rm PIT, \,
min}$ can be as low as $\approx 10^{-7}$ (hence
$\alpha_{\rm sat} \approx 10^{-6}$, see the footnote
\ref{bondarescu_ampl}) for fiducial values of the stellar
parameters $\nu=500$~Hz, $T=10^8$~K, and $R=10$~km (see
Eq.\ (16) or (35) of Ref.\ \cite{bw13}). However, this
result does not convince us because of the following
reasons. ($i$) The principal mode numbers of the daughter
modes in Ref.\ \cite{bw13} are $n_\mathrm D\sim 100$, but
their viscous damping times $\tau_D$ are just 50 times
smaller than the corresponding time $\tau_{\rm S \, 0}$ for
$m=2$ $r$-mode, although one would expect $\tau_D/\tau_{\rm
S \, 0} \sim 1/n_D^2=10^{-4}$ for $n_D=100$.
($ii$) The mutual friction dissipation was completely
ignored in Ref.\ \cite{bw13}, although it is an extremely
efficient damping mechanism for inertial modes in
superfluid NS matter \cite{yl03a}. If included, mutual
friction will increase dramatically the damping rates of
the daughter modes and hence increase the saturation
amplitude given by the lowest parametric instability
threshold in triplets of the $r$-mode and a couple of
inertial modes (see Eq.\ (4) of Ref.\ \cite{bw13}).
($iii$) To saturate $r$-mode at $\alpha\sim 10^{-6}$, the
amplitudes of the inertial daughter modes should reach the
value of $|C_\mathrm D|\sim 10^{-7}$, i.e. be of the same
order of magnitude as (or even larger) the amplitude of the
saturated $r$-mode (see Eq.\ (1) of Ref.\ \cite{btw07}).
However, according to the ``triangular'' selection rule for
the mode couplings (e.g., Ref.\ \cite{btw04a}), such
inertial modes can nonlinearly interact with plenty of
other oscillation modes and can easily find a mode triplet
with negligible detuning and relatively low ($<100$)
principal mode numbers of daughter modes. This will lead to
{\it lower} saturation amplitudes for inertial modes with
$n_\mathrm{D}\sim 100$ than for $r$-mode and thus will make
it impossible for these modes to saturate $r$-mode at
$\alpha\sim 10^{-6}$. \label{BW13} }

Summarizing, to the best of our knowledge, all attempts to
explain the significant number of rapidly rotating warm NSs
have been made under rather artificial assumptions that
either cannot be fully justified or even contradict the
up-to-date calculations available in the literature.

\section{Superfluid and normal modes}
\label{Sec_sflnorm}

\subsection{Two main assumptions}
\label{Subsec_2assum}

In this section we formulate and discuss two main assumptions
which are made in order to explain observations.

As it has been mentioned above, two types of inertial
modes, superfluid and normal ones, exist in rotating NSs.
Strictly speaking, these two types are clearly distinct
only if one sets to zero the so-called {\it coupling
parameter} $s$ \cite{gk11,gkcg13,kg13,gkgc14}. In the
absence of other mechanisms of mode decoupling (see the end
of this section), $s=s_{\rm EOS}$, where the parameter
$s_{\rm EOS}$ depends only on an EOS of superdense matter
and is given by \cite{gk11}
\begin{equation}
s_{\rm EOS} \equiv \frac{n_e}{n_b}\,
\frac{\partial P(n_b, \, n_e)/\partial n_e}{\partial P(n_b, \, n_e)/\partial n_b}.
\label{s}
\end{equation}
Here $n_b$ and $n_e$ are, respectively, the baryon and
electron number densities. As it was shown in Refs.\
\cite{gk11, kg13}, when $s$ vanishes, equations governing
superfluid and normal modes decouple into two independent
systems of equations. In this approximation, a system that
describes the normal modes can be written in exactly the
same form as for a nonsuperfluid star. Hence, the spectrum
and eigenfunctions of normal modes coincide with the
corresponding quantities of a normal star, and oscillation
frequencies $\omega$ are independent of NS temperature
$T^\infty$. Superfluid inertial modes, in turn, do not have
a counterpart in normal stars; unlike the normal modes,
$\omega$ for superfluid modes is a strong function of
$T^\infty$.

In reality, the actual coupling parameter $s$ is
small but finite (for example, for APR EOS $s_{\rm EOS}
\sim 0.01$--$0.03$ \cite{gk11}). This leads to a strong
interaction of normal and superfluid modes when their
frequencies become close to one another. As a result,
instead of crossings of these modes in the
$\omega-T^\infty$ plane, one has avoided crossings: As
$T^\infty$ varies, the superfluid mode turns into the
normal mode and vice versa.

\begin{figure*}
    \begin{center}
        \leavevmode
        \includegraphics[width=6.4in]{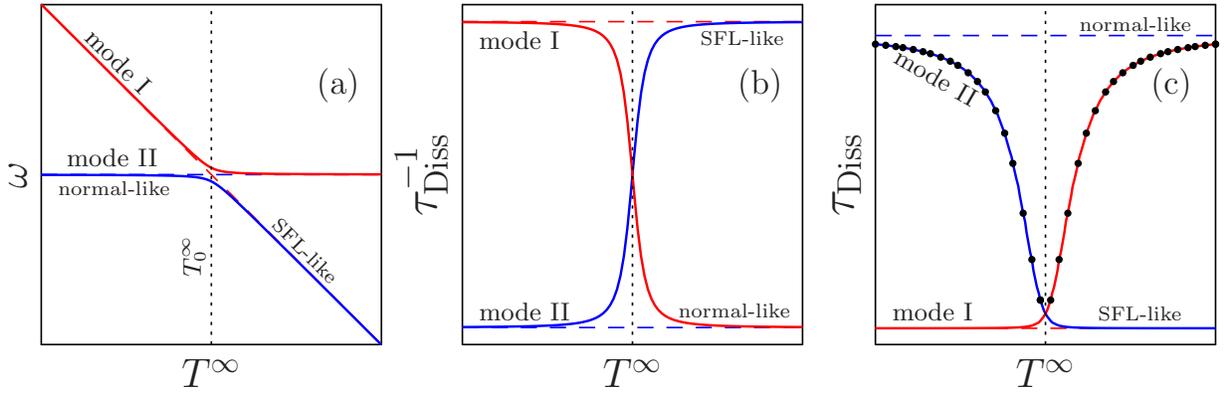}
    \end{center}
    \caption{(color online)
A schematic plot showing (a) oscillation frequencies, (b)
inverse damping time scale $\tau^{-1}_{\rm Diss}$, and (c)
$\tau_{\rm Diss}$ versus temperature $T^\infty$  for two
oscillation modes (I and II) of a superfluid NS, which
experience avoided crossing at $T^\infty= T_0^\infty$.
Dashes correspond to an approximation of independent
oscillation modes ($s=0$), solid lines are plotted for
exact solution allowing for the interaction of modes I and
II. The vertical dotted line indicates $T_0^\infty$. Filled
circles in panel (c) illustrate the results shown in Fig.\
12 of Ref.\ \cite{ly03}. See text for details.
    }
    \label{Fig_scheme}
\end{figure*}

These points are illustrated in Fig.\ \ref{Fig_scheme}(a)
where we (schematically) present oscillation frequency
$\omega$ as a function of $T^\infty$ for two neighboring
modes of a superfluid NS (these modes are denoted as ``I''
and ``II'', see the figure). At $T^\infty<T_0^\infty$, mode
I behaves itself as a superfluid one (that is, its
frequency depends on $T^\infty$), while mode II
demonstrates the normallike behavior. At $T^\infty \approx
T_{0}^\infty$, the frequencies of both modes come closer
and equations describing superfluid and normal modes become
strongly coupled. This results in an {\it avoided} crossing
of modes: At $T^\infty>T_0^\infty$  mode II starts to
behave as a superfluid mode while mode I becomes
normallike. In contrast, assuming $s=0$, one would obtain
crossing of modes instead of avoided crossing (see the
dashed lines in the figure); in that case superfluid and
normal modes would not ``feel'' each other.

The qualitative behavior of oscillation modes in superfluid
NSs described above has been confirmed by direct
calculation of radial \cite{kg11,gkcg13} and nonradial
\cite{gkgc14} oscillation modes. The concept of weakly
interacting superfluid and normal modes has also been used
in Refs.\ \cite{cg11, gkcg13} for a detailed analysis of
nonradial oscillation spectra of nonrotating NSs and
damping of these oscillations.

Unfortunately, self-consistent calculations of
oscillations of rotating superfluid NSs at finite temperatures
are still unavailable in the literature.
However, it seems natural that
the behavior of inertial modes
(in particular, $r$-modes) in superfluid NSs
should be quite similar.
The results of Refs.\ \cite{lm00, yl03a, yl03b, ly03}
provide indirect independent confirmation
of this assumption (see below).

Thus, our first main assumption is

1. {\it An oscillation mode of a superfluid rotating NS,
which behaves, at some $T^\infty$, as a normal quadrupole
$m=2$ $r$-mode ($r^o$-mode) can, as the temperature
gradually changes, transform into a superfluidlike inertial
mode ($i^s$-mode).}

Our second main assumption is

2. {\it Dissipative damping of a NS oscillation mode in the
regime when it mimics the $m=2$ $r^o$-mode is much smaller
than damping of this mode in the superfluid-like
($i^s$-mode) regime} [see Figs.\
\ref{Fig_scheme}(b)--\ref{Fig_scheme}(c), which show a
qualitative dependence of the damping time scale $\tau_{\rm
Diss}$ and its inverse $\tau_{\rm Diss}^{-1}$ on $T^\infty$
for the same two modes as in Fig.\ \ref{Fig_scheme}(a)].

What is the second assumption based on?

First, it is based on the analysis of $\tau_{\rm Diss}$ for
nonradial oscillations of a nonrotating NS
\cite{gkcg13,gkgc14}. As it was demonstrated in Ref.\
\cite{gkcg13}, damping of oscillation modes due to the
shear viscosity in the superfluidlike regime occurs
approximately ten times faster than their damping in the
normallike regime. The reasons for that are discussed in
detail in Sec.\ 7.4 of Ref.\ \cite{gkcg13} and should be
applicable to $r$-modes. This is also in line with the
results of Refs.\ \cite{ly03, yl03a}, where it was found
that $\tau_{\rm S}$ for the zero-temperature $i^s$-modes is
generally more than 1 order of magnitude smaller than for
normal $r^o$-modes (compare Table 1 of Ref.\ \cite{ly03}
and Table 2 of Ref.\ \cite{yl03a}).

But the main dissipation mechanism, which leads to a
drastic difference (by orders of magnitude) of $\tau_{\rm
Diss}$ in superfluid- and normallike regimes, is the mutual
friction between the superfluid and normal matter
components \cite{als84, mendell91b, asc06}. The friction
occurs because of electron scattering off the magnetic
field of Feynman-Onsager vortices. The corresponding
magnetic field is generated because of entrainment
\cite{ab75} of superconducting protons by the motion of
superfluid neutrons.

This mechanism tends to equalize the velocities of normal
and superfluid components; it does not noticeably affect
dissipation of the normal modes, since for normal modes
these velocities approximately coincide (comoving motion).
On the opposite, mutual friction is extremely effective for
superfluid modes, because in that case the difference
between the normal and superfluid velocities is large
(countermoving motion). In application to $r$-modes the
effects of mutual friction were studied in detail in Refs.\
\cite{lm00, ly03, agh09, hap09}. In particular, the damping
time scale for {\it normal} $r$-modes ($r^o$-modes) due to
mutual friction was shown to be
\begin{equation}
\frac{1}{\tau^{\rm norm}_{\rm MF}}=\frac{1}{\tau^{\rm norm}_{{\rm MF} \, 0}} \, \left( \frac{\Omega}{\Omega_0}\right)^5,
\label{nu_MFnorm}
\end{equation}
where $\tau_{{\rm MF} \, 0}^{\rm norm} \sim 10^3$--$10^4$~s
\cite{lm00,ly03}.
Superfluid $r$-modes ($r^s$-modes) and superfluid inertial
modes ($i^s$-modes) were studied, for the first time, in
Refs.\ \cite{ly03} and \cite{yl03a}, respectively; for the
damping time scale of these modes due to mutual friction
they obtain
\begin{equation}
\frac{1}{\tau^{\rm sfl}_{\rm MF}}
=\frac{1}{\tau^{\rm sfl}_{{\rm MF} \, 0}} \, \frac{\Omega}{\Omega_0},
\label{nu_MFsfl}
\end{equation}
where $\tau^{\rm sfl}_{{\rm MF}\, 0} \sim 0.1$~s (see Table
1 in Ref.\ \cite{ly03} and Table 2 in Ref.\ \cite{yl03a}).
It is interesting that $i^s$-modes were also presumably
found in Ref.\ \cite{lm00} (see the resonances in their
Fig.\ 6 and the corresponding discussion in that
reference).

The results obtained by Lee and Yoshida \cite{ly03,yl03a}
indirectly confirm our main assumptions 1 and 2. These
authors employed the zero temperature approximation
($T^\infty=0$) and varied the so-called ``entrainment''
parameter $\widetilde{\eta}$ ($\eta$ in their paper), that
parametrizes interaction between the superfluid neutrons
and superconducting protons. It follows from the
microphysics calculations \cite{gh05, gkh09b, gusakov10}
that $\widetilde{\eta}$ is a function of $T^\infty$. Hence,
its variation is {\it analogous} to a variation of stellar
temperature. In other words, the eigenfrequencies and
eigenfunctions for the superfluid oscillation modes should
depend on $\widetilde{\eta}$, while these for the normal
modes should be almost insensitive to this parameter. Thus,
all the peculiarities in the behavior of oscillation modes
with changing $T^\infty$ discussed above should also be
observed in calculations of Refs.\ \cite{ly03,yl03a}, where
$\widetilde{\eta}$ is varied. (In particular, Fig.\
\ref{Fig_scheme} should still be applicable, provided that
one replaces $T^\infty$ with $\widetilde{\eta}$ there.)

And indeed, Lee and Yoshida \cite{ly03, yl03a} found
numerous avoided crossings of superfluid and normal
inertial modes (see their Figs.\ 5--8 in Ref.\
\cite{yl03a}). Concerning $r$-modes, in Ref.\ \cite{ly03}
they found avoided crossing between the $m=2$ $r^s$-mode
and one of the normal inertial $i^o$-modes (see their Fig.\
7) and {\it crossings} of the $m=2$ $r^o$-mode with two
superfluid inertial modes (see their Fig.\ 8). In the
latter case, Lee and Yoshida emphasized on p.\ 409 that
``it is quite difficult to numerically discern whether the
mode crossings result in avoided crossings or degeneracy of
the mode frequencies at the crossing point.'' If our
interpretation is correct, there should be avoided
crossings.

This point of view is supported by Fig.\ 12 of the same
Ref.\ \cite{ly03}. The figure shows the time scale
$\tau_{{\rm MF}\, 0}$ [corresponding to our time $\tau_{\rm
MF \, 0}^{\rm norm}$, introduced in Eq.\ (\ref{nu_MFnorm})]
for the $m=2$ $r^o$-mode as a function of
$\widetilde{\eta}$ for the same stellar parameters as in
Fig.\ 8 of that reference. One can see that $\tau_{{\rm
MF}\, 0}$ in Fig.\ 12 sharply decreases (by a few orders of
magnitude) at the values of $\widetilde{\eta}$ at which one
observes crossing of the $r^o$- and $i^s$-modes in Fig.\ 8.
This is exactly what one would expect if our assumptions 1
and 2 are correct. Near the crossing of modes (which is
avoided crossing in reality) the $m=2$ $r^o$-mode starts to
transform into the $i^s$-mode, and hence $\tau_{{\rm MF}\,
0}$ drops down rapidly. Moving away from the avoided
crossing (by decreasing or increasing $\widetilde{\eta}$)
the solution found by Lee and Yoshida resembles more and
more the $m=2$ $r^o$-mode. Consequently, $\tau_{{\rm MF}\,
0}$ grows on both sides of the resonance, approaching the
asymptote value corresponding to the pure (with no
admixture) $m=2$ $r^o$-mode. The results obtained in Fig.\
12 of Ref.\ \cite{ly03} are shown qualitatively by filled
circles in our Fig.\ \ref{Fig_scheme}(c).

The fact that Lee and Yoshida \cite{ly03} fail to
discriminate between crossing and avoided crossing of modes
in their Fig.\ 8 indicates that the real coupling parameter
$s$ responsible for the interaction of $m=2$ $r^o$- and
$i^{s}$-modes is actually much smaller than the parameter
$s_{\rm EOS}$ given by Eq.\ (\ref{s}). The reason is the
stellar matter only weakly deviates from the
beta-equilibrium state in the course of the $m=2$
$r^o$-mode oscillations [the deviation $\delta \mu \sim
(\Omega/\Omega_0)^4$ \cite{lm00} is small since $\Omega \ll
\Omega_0$]. It can be shown \cite{gk11,gkcg13, kg13,gkgc14}
that in that case the superfluid degrees of freedom
decouple from the normal ones especially well. According to
our preliminary estimates, the real coupling parameter can
be of the order of $s \sim s_{\rm EOS} \,
(\Omega/\Omega_0)^2$. If this estimate is correct then for
$s_{\rm EOS}=0.01$ and $\Omega/\Omega_0 =0.1$ one has $s
\sim 10^{-4}$. However, in view of the existing
uncertainties, in this paper we adopt the larger value, $s
= 0.001$. We checked that the variation of $s$ within the
very wide range (by orders of magnitude) does not affect
our principal results.

\subsection{Mixing the modes}
\label{Subsec_mix}

Obviously the fact that the real oscillation modes of
superfluid NSs demonstrate, depending on $T^\infty$, either
normal- or superfluidlike behavior should have a major
effect on the stability region discussed in Sec.\
\ref{Subsec_obs_stab}. To describe this effect it is
necessary to understand how the time scales $\tau_{\rm S}$,
$\tau_{\rm MF}$, and $\tau_{\rm GR}$ are modified during
the transformation of the mode from the normallike to
superfluidlike regime (see Fig.\ \ref{Fig_scheme}). Since
there are no accurate calculations of these time scales in
the literature, below we develop a simple phenomenological
model evoked by the perturbation theory of quantum
mechanics.

Assume for a moment that the coupling parameter $s=0$, so
that the systems of equations describing the superfluid and
normal oscillation modes are completely decoupled. The
solution to these systems of equations describes two types
of independent modes, the superfluid and normal ones. Let
us present the eigenfunctions of normal modes in the form
of a column vector $\Psi_{\rm norm}$ and those of
superfluid modes in the form of a column vector $\Psi_{\rm
sfl}$. Assume further that $\Psi_{\rm norm}$ and $\Psi_{\rm
sfl}$ are normalized by the one and the same oscillation
energy $E_{\rm c}$ and that the time scale $\tau_{X}$ of
damping/excitation of oscillations due to some dissipation
mechanism [e.g., shear viscosity ($X={\rm S}$), mutual
friction ($X={\rm MF}$), or gravitational radiation
($X={\rm GR}$)] is given by the general formula of the form
\begin{equation}
\frac{1}{\tau_{ X}} = -\frac{1}{2 E_{\rm c}} \,
\frac{dE_{\rm c}}{dt} =-\frac{1}{2 E_{\rm c}} \, (\Psi, \,
\hat{A} \, \Psi), \label{tau_phen}
\end{equation}
where $\hat{A}$ is a matrix differential operator and
$(\Psi_1, \, \Psi_2)$ is a scalar product, both specified
by the actual mechanism of dissipation. For example, for
$X={\rm S}$ or MF the scalar product
is defined as (e.g., Ref.\ \cite{yl03a})%
\footnote{The definition of scalar product for $X={\rm GR}$
follows, e.g., from Eqs.\ (36) and (37) of Ref.\
\cite{yl03a}.}
%
\begin{equation}
(\Psi_1, \Psi_2) \equiv \int_{\rm star} \Psi_1^\dagger \, \Psi_2 \, dV,
\label{defmult}
\end{equation}
where the integration is performed over the NS volume $V$.
To determine the time scale $\tau_{X}^{\rm norm}$ for
normal modes one should set $\Psi \equiv \Psi_{\rm norm}$
in Eq.\ (\ref{tau_phen}); similarly, to determine the time
scale $\tau_{X}^{\rm sfl}$ for superfluid modes one should
assume $\Psi \equiv \Psi_{\rm sfl}$. Note that for normal
$r$-modes the time scales $\tau_{\rm GR}^{\rm norm}$ and
$\tau_{\rm S}^{\rm norm}$ have been already calculated in
Sec.\ \ref{Sec_input} and are given by, respectively, Eqs.\
(\ref{tauGR2}) and (\ref{tauS2}).

As has been mentioned above, in reality the parameter $s$
is small but finite. This means that the eigenfunctions
$\Psi_{\rm norm}$ and $\Psi_{\rm sfl}$ approximate well the
exact solution {\it far from the avoided crossings of
neighboring modes} ($\Psi_{\rm norm}$ describes well the
exact solution in the normallike regime, while $\Psi_{\rm
sfl}$ does so in the superfluidlike regime). However, in
the vicinity of an avoided crossing the eigenfunctions of
the exact solution should be presented as a linear
superposition of $\Psi_{\rm norm}$ and $\Psi_{\rm sfl}$. In
particular, in Fig.\ \ref{Fig_scheme} avoided crossing
occurs between modes I and II. Denoting the corresponding
eigenfunctions as $\Psi_{\rm I}$ and $\Psi_{\rm II}$, one
can write
\begin{eqnarray}
\Psi_{\rm I} &=& -{\rm sin} \theta(x) \, \Psi_{\rm norm} +
{\cos} \theta(x) \, \Psi_{\rm sfl},
\label{PsiReal1}\\
\Psi_{\rm II} &=& ~~\, {\rm cos} \theta(x) \, \Psi_{\rm
norm} + {\sin} \theta(x) \, \Psi_{\rm sfl},
\label{PsiReal2}
\end{eqnarray}
where ${\cos} \theta(x)$ and ${\sin}\theta(x)$ guarantee
the correct normalization of the eigenfunctions $\Psi_{\rm
I}$ and $\Psi_{\rm II}$ by the oscillation energy $E_{\rm
c}$, while the function $\theta(x)$ determines how the
normal mode transforms into the superfluid one (and vice
versa). This function depends on the parameter $x \equiv
(T^\infty-T_0^\infty)/\Delta T^\infty$ [see Fig.\
\ref{Fig_scheme}(a)] and ranges from $0$ to $1$ on a
temperature scale specified by the characteristic width
$\Delta T^\infty$ of the avoided crossing, $\Delta T^\infty
\sim s \, T^\infty_0$. The exact form of the function
$\theta(x)$ can be found only by direct solution to the
coupled oscillation equations. However, using as the
analogy the problem of intersection of electron terms in
molecules (see, e.g., Ref.\ \cite{ll77}, Sec.\ 79), one can
immediately write down an approximate expression for
$\theta(x)$ that correctly reproduces its main properties,
\begin{equation}
\theta(x) = \frac{1}{2} \left[ \frac{\pi}{2}+ \arctan (x)
\right]. \label{theta}
\end{equation}
Consider, for example,  mode II. At $x \rightarrow -\infty$
one has $\theta(x) \rightarrow 0$, and it follows from Eq.\
(\ref{PsiReal2}) that  mode II is in the normallike regime
($\Psi_{\rm II}=\Psi_{\rm norm}$); at $x \rightarrow
+\infty$ one obtains $\theta(x) \rightarrow \pi/2$, which
corresponds to superfluidlike behavior of mode II
($\Psi_{\rm II}=\Psi_{\rm sfl}$).

Now, substituting Eqs.\ (\ref{PsiReal1}) and (\ref{PsiReal2}) into (\ref{tau_phen})
and neglecting the interferential terms of the form%
\footnote{The contribution of these terms can be neglected
since the time scales $\tau_{X}^{\rm norm}$ and
$\tau_{X}^{\rm sfl}$ differ by at least 1 order of
magnitude (see Sec.\ \ref{Subsec_wind} for details).}
%
\begin{equation}
 -\frac{1}{2 E_{\rm c}} \, \,\,
 {\cos}\theta(x)\, {\sin}\theta(x) \, (\Psi_{\rm norm}, \, \hat{A} \, \Psi_{\rm sfl}),
\label{interfere}
\end{equation}
one gets
\begin{equation}
   \frac{1}{\tau_{X}} \approx
    \frac{1}{\tau_{X}^{\rm norm}} \, {\sin}^2 \theta(x)
   + \frac{1}{\tau_{X}^{\rm sfl}} \, {\cos}^2 \theta(x) \label{tau_resA}
\end{equation}
for mode I and
\begin{equation}
  \frac{1}{\tau_{X}} \approx \frac{1}{\tau_{X}^{\rm norm}} \, {\cos}^2 \theta(x)+
  \frac{1}{\tau_{X}^{\rm sfl}} \, {\sin}^2 \theta(x)
\label{tau_resB}
\end{equation}
for mode II. These are the main formulas of our approximate
model. Their use for $X={\rm S, \, MF,\, GR}$ enables us to
plot the instability windows for the real oscillation modes
(similar to modes I and II shown in Fig.\
\ref{Fig_scheme}).

\section{Realistic instability windows and three-mode regime}
\label{Sec_wind_regime}

\subsection{Realistic instability windows}
\label{Subsec_wind}

Let us assume that a certain oscillation mode of a rotating
superfluid NS (by analogy with the previous section we will
refer to it as mode II) behaves like the $m=2$ $r^o$-mode
at low temperatures, and that at $T^\infty=T^\infty_0$ it
experiences an avoided crossing with another mode (with the
same $m=2$, let us call it mode I), which behaves like a
superfluid inertial mode ($i^s$-mode) at low $T^\infty$
(exactly as in the scheme in Fig.\ \ref{Fig_scheme}). After
avoided crossing, mode I starts to behave as an $m=2$
$r^o$-mode, and mode II as an $i^s$-mode. Let us determine
the instability windows for these modes.

The instability windows are defined by the following
inequality (see also Sec.\ \ref{Subsec_obs_stab} above):
\begin{equation}
\frac{1}{\tau_{\rm GR}}+\frac{1}{\tau_{\rm
S}}+\frac{1}{\tau_{\rm MF}}<0.
\label{instab}
\end{equation}
Each of these times cales can be calculated using Eq.\
(\ref{tau_resB}) for mode II and Eq.\ (\ref{tau_resA}) for
mode I. One only needs to specify the values for $\tau_{
X}^{\rm norm}$ and $\tau_{X}^{\rm sfl}$, which will be
employed in each case.

($i$) Shear viscosity ($ X={\rm S}$). The damping time
scale $\tau_{\rm S}^{\rm norm}$ for the $m=2$ $r^o$-mode is
determined by Eq.\ (\ref{tauS2}). According to the
discussion in Sec.\ \ref{Subsec_2assum}, $\tau_{\rm S}^{\rm
sfl}$ for the $i^s$-mode is taken to be
\begin{equation}
\tau_{\rm S}^{\rm sfl}=c_{\rm S} \, \tau_{\rm S}^{\rm norm},
\label{tauSsfl}
\end{equation}
where $c_{\rm S}=0.1$. Since the mutual friction
dissipation dominates for the superfluid $i^s$-mode [see
item ($ii$) below and compare Eqs.\ (\ref{nu_MFsfl}) and
(\ref{tauSsfl})], the specific value of the coefficient
$c_{\rm S}$ is not important for our scenario; one can take
$1$ or $0.01$ instead of $0.1$, and the main results will
not change.

($ii$) Mutual friction ($X={\rm MF}$). The damping time
scale $\tau_{\rm MF}^{\rm norm}$ is given by Eq.\
(\ref{nu_MFnorm}) with $\tau_{\rm MF \, 0}^{\rm
norm}=10^4$~s; the time $\tau_{\rm MF}^{\rm sfl}$ is
determined from Eq.\ (\ref{nu_MFsfl}) with $\tau_{\rm MF \,
0}^{\rm sfl}=2.5$~s. Our scenario is insensitive to the
actual choice of $\tau_{\rm MF \, 0}^{\rm norm}$ because
the mutual friction is not a dominating dissipative process
for normal modes. However, it is crucial that $\tau_{\rm MF
\, 0}^{\rm sfl}$ be sufficiently small, $\tau_{\rm MF \,
0}^{\rm sfl} \lesssim 100$~s.

\begin{figure*}
    \begin{center}
        \leavevmode
        \includegraphics[width=6.5in]{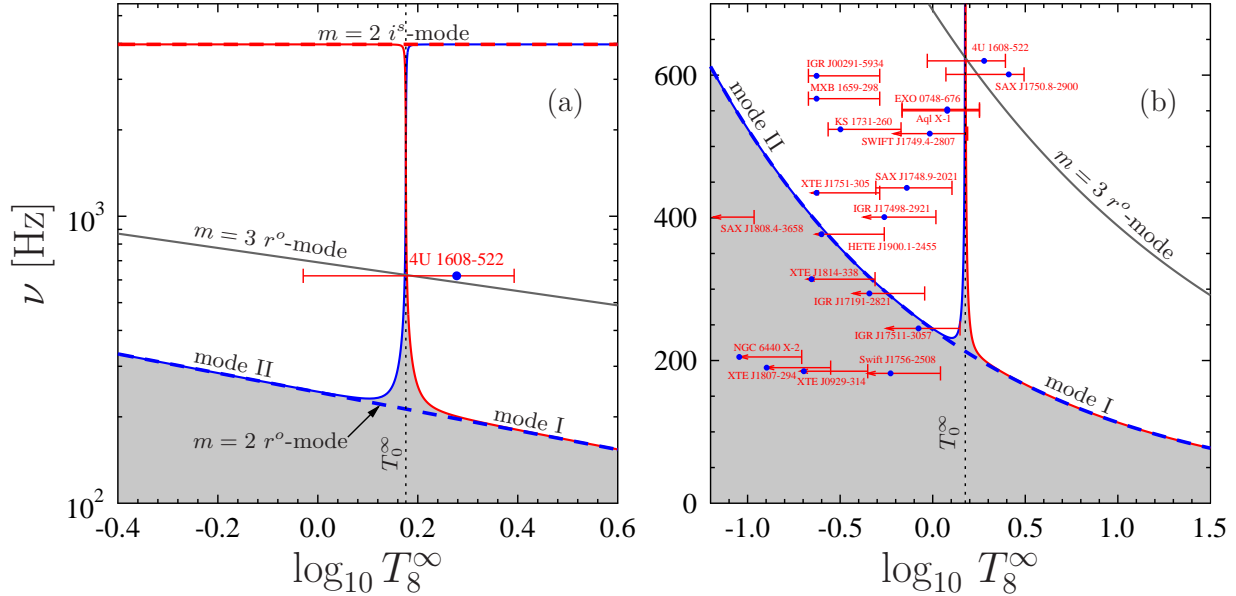}
    \end{center}
    \caption{(color online)
Instability curves for superfluid NS oscillations. The
solid curves correspond to $m=2$ modes I and II (red and
blue online, respectively), which experience avoided
crossing at $T_0^\infty=1.5 \times 10^8$~K. The coupling
parameter was chosen to be $s=0.001$. The dashed  curves
correspond to the $m=2$ $r^o$- and $i^s$-modes (blue and
red online, respectively) plotted under the assumption that
they are completely decoupled ($s=0$). The grey line is the
instability curve for the $m=3$ $r^o$-mode, plotted
ignoring the resonance coupling with the superfluid modes.
The temperature $T_0^\infty$ is shown by the vertical
dotted line. Similar to Fig.\ \ref{Fig_typical_obs}, the
panel (b) shows temperatures and frequencies of the sources
from Table \ref{Tab_LMXB_Observ}. Only the fastest source
4U~1608-522 is shown in the panel (a). See text for
details.
    }
    \label{Fig_instab}
\end{figure*}

($iii$) Gravitational radiation ($X={\rm GR}$). The time
scale $\tau_{\rm GR}^{\rm norm}$ is given by Eq.\
(\ref{tauGR2}); $\tau_{\rm GR}^{\rm sfl}$ is taken to be
\begin{equation}
\tau_{\rm GR}^{\rm sfl} = c_{\rm GR} \, \tau_{\rm GR}^{\rm
norm}, \label{tauGRsfl}
\end{equation}
where $c_{\rm GR}=100$. Such an expression for the
gravitational radiation time scale for the $i^s$-mode
agrees qualitatively with the results of Refs.\
\cite{ly03,yl03a} [see Eq.\ (44) and Table 2 of Ref.\
\cite{yl03a}], where even longer time scales were obtained,
corresponding to $c_{\rm
GR}>10^4$
(see also \cite{gkgc14}). For readability of Fig.\
\ref{Fig_instab}(a) we take $c_{\rm GR}=100$, thus
underestimating
$\tau_{\rm GR}^{\rm sfl}$ for the $m=2$ $i^s$-mode
significantly. Increasing of $c_{\rm GR}$ (and even further
decreasing of $c_{\rm GR}$ down to $\sim 1$) does not
affect the scenario suggested in this paper.

Instability curves for modes I (solid line; red online) and
II (solid line; blue online) are shown in Figs.\
\ref{Fig_instab}(a)--\ref{Fig_instab}(b). The curves are
obtained by making use of Eqs.\
(\ref{tau_resA})--(\ref{tauGRsfl}) with the coupling
parameter $s=0.001$. Panel (b) is a version of panel (a),
but plotted in a different scale. The dotted line in Figs.\
\ref{Fig_instab}(a)--\ref{Fig_instab}(b) corresponds to the
temperature $T^\infty_0=1.5 \times 10^8$~K, at which the
modes I and II experience avoided crossing. In addition,
Figs.\ \ref{Fig_instab}(a)--\ref{Fig_instab}(b) show the
instability curves for ($i$) octupole $m=3$ $r^o$-mode
(grey solid line; to plot it, we take the characteristic
time scales $\tau_{\rm S}$ and $\tau_{\rm GR}$ from Sec.\
\ref{Sec_input} and ignore the mutual friction, $\tau_{\rm
MF} \equiv \infty$); ($ii$) $m=2$ $r^o$-mode (dashed line;
blue online); ($iii$) superfluid $i^{s}$-mode with $m=2$
(dashed line; red online). The latter curves ($i$)--($iii$)
are obtained using the approximation $s=0$ (neglecting the
interaction between the superfluid and normal modes).

As one would expect, far from the avoided crossing point
the solid (modes I and II) and dashed ($r^o$ and
$i^s$-modes) lines almost coincide. The region where $m=2$
modes I, II, and the octupole $m=3$ $r^o$-mode are
simultaneously stable is filled with grey in Figs.\
\ref{Fig_instab}(a)--\ref{Fig_instab}(b). The presence of
the ``stability peak'' at $T^\infty \approx T_{0}^\infty$
is an important characteristic feature of this region. The
height of the peak is determined by the lowest-frequency
intersection of the mode II instability curve with the
other instability curves. The instability curves for modes
I and II intersect at a very high frequency $\nu \approx
1580$~Hz; hence, the lowest-frequency intersection
corresponds to that with the octupole $m=3$ $r^o$-mode and
occurs at $\nu \approx 625$~Hz. As a result, at
$T^\infty=T_0^\infty$ the most unstable mode is the $m=3$
$r^o$-mode,
and the height of the stability peak is $\nu \approx 625$~Hz.%
\footnote{The octupole $m=3$ $r^o$-mode can also experience
a resonant coupling with the superfluid $m=3$ oscillation
modes. However, the correspondent resonance temperatures
are unlikely to be close to those for the $m=2$ $r^o$-mode.
Therefore, at $T^\infty \approx T_0^\infty$ the instability
curve for the $m=3$ $r^o$-mode will hardly be essentially
affected by coupling with superfluid modes.}

As follows from Fig.\ \ref{Fig_instab}, the evolution of a
NS with such a complicated structure of instability windows
can be accompanied by excitation of each of the three
oscillation modes. Therefore, prior to discussing the
evolution tracks one should formulate the equations
describing an oscillating star in a three-mode regime.

\subsection{Three-mode regime}
\label{Subsec_regime}

The equations governing the evolution of a NS and allowing
for possible excitation of the three modes (I, II, and
$m=3$ $r^o$-mode) can be derived in much the same fashion
as it was done in Sec.\ \ref{Sec_input} [see the one-mode
equations (\ref{thermal}), (\ref{dadt}), and
(\ref{dOmegadt}) in that section]. If all the modes are
nonsaturated, they can be written as
\begin{eqnarray}
\frac{d \alpha_i}{dt} &=& -\alpha_i \left(
\frac{1}{\tau_{{\rm GR}\, i}}+\frac{1}{\tau_{{\rm Diss}\,
i}} \right),
\label{alpha_i}\\
\frac{d \Omega}{dt} &=& - \sum_i \frac{2 \, Q_i \,
\alpha_i^2 \, \Omega}{\tau_{{\rm Diss}\,
i}}+\dot{\Omega}_{\rm acc},
\label{Omega_i}\\
C_{\rm tot} \frac{dT^\infty}{dt} &=& \sum_i W_{{\rm Diss}\,
i}-L_{\rm cool} + K_{\rm n} \dot{M} c^2, \label{therm_i}
\end{eqnarray}
where we neglect the terms $\propto \alpha_{i}^3$.
The index $i$ in Eqs.\ (\ref{alpha_i})--(\ref{therm_i})
runs over the mode types, and
\begin{eqnarray}
W_{{\rm Diss}\, i} &=& \frac{2 E_{{\rm c}\, i}}{\tau_{{\rm
Diss}\, i}},
\label{Wdiss_i}\\
\frac{1}{\tau_{{\rm Diss}\, i}} &=& \frac{1}{\tau_{{\rm
S}\, i}}+\frac{1}{\tau_{{\rm MF}\, i}}, \label{tauDiss_i}
\end{eqnarray}
where $\tau_{{\rm S}\, i}$ and $\tau_{{\rm MF}\, i}$ for
modes I and II are calculated as it is described in Sec.\
\ref{Subsec_wind}, while for the octupole $m=3$ $r^o$-mode
they are calculated as described in Sec. \ref{Sec_input}
(we neglect the effects of mutual friction on damping of
the octupole $r^o$-mode).

Thus, only the quantities $E_{{\rm c}\, i}$ and $Q_i$ in
Eqs.\ (\ref{Omega_i}) and (\ref{Wdiss_i}) are left to be
determined. The corresponding Eqs.\ (\ref{Ec}) and
(\ref{Q}) for the octupole $r^o$-mode are presented in
Sec.\ \ref{Sec_input}. In the case of modes I and II one
can argue as follows. First, let us discuss mode II. At low
$T^\infty$ (before the avoided crossing), it behaves like
the $m=2$ $r^o$-mode. Accordingly, its canonical angular
momentum $J_{{\rm c}\, {\rm II}}$ is given by Eq.\
(\ref{Jc2}), where the coefficient $\widetilde{J} \approx
1.6353 \times 10^{-2}$. At the avoided crossing point the
behavior of the mode changes and it turns into the
$i^s$-mode. However, since the canonical angular momentum
is an adiabatic invariant \cite{fs78a,hl00,wagoner02},
$J_{{\rm c} \, {\rm II}}$ is conserved (neglecting
dissipative processes) and stays the same even after
passing the avoided crossing. Without any loss of
generality, one can assume it to be still related to the
oscillation amplitude $\alpha_{\rm II}$ by exactly the same
Eq.\ (\ref{Jc2}) (with the same $\widetilde{J}=1.6353
\times 10^{-2}$), as before the avoided crossing. This
assumption, which should be treated as the definition of
the amplitude $\alpha_{\rm II}$ in the superfluidlike
regime, has already been implicitly employed when deriving
the system of Eqs.\ (\ref{alpha_i})--(\ref{therm_i}). It
ensures that $\alpha_{\rm II}$ is continuous throughout the
avoided crossing region.

The same reasoning also holds true for mode I. For a given
$J_{{\rm c}\,i}$ the quantities $Q_i$ and $E_{{\rm c}\, i}$
can be found from Eqs.\ (\ref{Ec}) and (\ref{Q}). The
problem, however, consists in that the mode energy $E_{{\rm
c}\, i}$ depends on the oscillation frequency $\omega$,
which is only known for modes I and II in the normallike
regime [in that case, it is given by Eq.\ (\ref{o})]. In
the superfluidlike regime, $\omega$ depends not only on
$\Omega$, but also on $T^\infty$; unfortunately, the
function $\omega(\Omega,\, T^\infty)$ has not yet been
calculated. Below, for simplicity, we assume that the
frequency $\omega$ is determined by the same Eq.\ (\ref{o})
even in the superfluidlike regime. This assumption does not
influence our main conclusions and is well justified
because the range of $T^\infty$, which is of interest in
our scenario (see Sec.\ \ref{Sec_climb}), is located near
avoided crossings of modes. In that region $\omega$ for
both modes can indeed be estimated from Eq.\ (\ref{o}).
Beyond this region any mode in the superfluidlike regime is
stable, unexcited, and, correspondingly, not important for
NS evolution.

Equations (\ref{alpha_i})--(\ref{therm_i}) are satisfied
if the oscillation amplitudes $\alpha_i$
are less than the correspondent
saturation amplitudes $\alpha_{{\rm sat}\, i}$.
In the following,
the saturation amplitudes for {\it all} the modes
are taken to be $\alpha_{{\rm sat}\, i}=10^{-4}$ .
Note that our main results are insensitive to the actual value
of $\alpha_{{\rm sat}\, i}$.%
\footnote{In particular, the choice of $\alpha_{{\rm sat}}$
for the $m=3$ $r^o$-mode appears to be insignificant and
does not even affect the position of the Cooling=Heating
curve (see Sec.\ \ref{Sec_climb}).}
If one or more modes are saturated, the evolution equations
can be derived in a similar way as it was done in Sec.\
\ref{Sec_input}.

\section{Our resonance uplift scenario}
\label{Sec_climb}

\begin{figure}
    \begin{center}
        \leavevmode
        \includegraphics[width=3.4in]{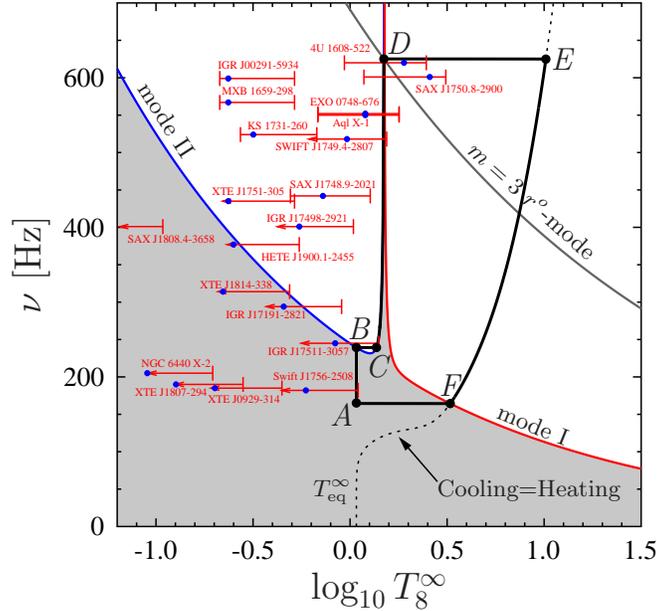}
    \end{center}
    \caption{(color online)
Evolution of the spin frequency $\nu$ and temperature
$T^\infty_8$ for a superfluid NS in LMXB allowing for the
avoided crossing of $m=2$ modes I and II. The corresponding
track $A$--$B$--$C$--$D$--$E$--$F$--$A$ is shown by the
thick solid line. The dotted line shows the Cooling=Heating
curve (see text for details). Other notations are the same
as in Fig.\ \ref{Fig_instab}.
    }
    \label{Fig_scenario}
\end{figure}

Using the results of the preceding sections,
we can examine quantitatively
how the resonance coupling of superfluid and normal modes
modifies the standard scenario discussed in Sec.\ \ref{Subsec_obs_stab}
(see also Fig.\ \ref{Fig_typical_obs}).

A typical NS evolution track
$A$--$B$--$C$--$D$--$E$--$F$--$A$ is shown in Fig.
\ref{Fig_scenario} by the thick solid line, calculated for
exactly the same model as the instability curves in Sec.\
\ref{Subsec_wind} (see Fig.\ \ref{Fig_instab}). Other
notations coincide with those in Fig.\ \ref{Fig_instab}. As
in Sec.\ \ref{Subsec_wind}, we suppose that mode I
experiences an avoided crossing with mode II at
$T^\infty=T_0^\infty=1.5 \times 10^8$~K.

To plot the Cooling=Heating curve (shown by the dotted line
in Fig.\ \ref{Fig_scenario}), we use Eq.\ (\ref{therm_i})
with $dT^\infty/dt=0$. When doing this we assume that all
the modes, which are unstable at a given temperature and
frequency, are saturated, while the stable modes have
vanishing oscillation amplitudes. This means that in each
point of the Cooling=Heating curve the neutrino luminosity
is exactly compensated by the stellar heating due to
nonlinear damping of {\it saturated} modes. Let us note
that in the stability region (the grey-filled area in the
figure) we do not use this definition, but instead, by
analogy with Fig.\ \ref{Fig_typical_obs}, continue the
Cooling=Heating curve
according to Eq.\ (\ref{Coolheat}).%
\footnote{The point is that the Cooling=Heating curve in
the instability region is almost indistinguishable from the
curve given by Eq.\ (\ref{Coolheat}); see the following
discussion herein.}
A break of the Cooling=Heating curve
at the intersection point with the instability curve for
the $m=3$ $r^o$-mode is imperceptible, because the
contribution of the octupole mode to stellar heating can be
neglected owing to a longer gravitational radiation time
scale for this mode [see Eq.\ (\ref{tauGR2})]. Therefore,
along the whole Cooling=Heating curve, the nonlinear
damping of mode I, behaving as the saturated $m=2$
$r^o$-mode, is the dominating heating mechanism. This means
that the Cooling=Heating curve, obtained while allowing for
the resonance coupling of modes, is practically {\it
indistinguishable} from that given by Eq.\ (\ref{Coolheat})
(see Sec.\ \ref{Subsec_obs_stab} and Fig.\ \ref{Fig_typical_obs}).%
\footnote{Due to this fact, it is easy to understand an
impact that the instability curve for $m=2$ $r^o$-mode has
on the stellar evolution track
$A$--$B$--$C$--$D$--$E$--$F$--$A$ (see its description in
the text). Point $F$ is determined by the intersection of
the Cooling=Heating curve, given by Eq.\ (\ref{Coolheat}),
with the instability curve; its frequency fixes the
frequency of point $A$. Point $B$ lies on the instability
curve at $T^\infty=T^\infty_\mathrm{eq}$, and specifies the
frequency of point $C$. Points $D$ and $E$ do not depend on
the position of $m=2$ $r^o$-mode instability curve.}

During the $A$--$B$ stage, a NS stays inside the stability
region and gradually spins up by accretion. This stage is
completely analogous to the $A$--$B$ stage of the standard
scenario shown in Fig.\ \ref{Fig_typical_obs}. At point
$B$, the star becomes unstable with respect to excitation
of mode II, which behaves there as the $m=2$ $r^o$-mode. In
the next stage $B$--$C$ the amplitude of mode II increases
and rapidly reaches saturation ($\alpha_{\rm
sat}=10^{-4}$). After that, the star heats up without any
significant variation of the spin frequency $\nu$. This
stage ends by reaching the stability peak at point $C$.

The next stage $C$--$D$ is the most interesting and is
absent in the standard scenario described in Sec.\
\ref{Subsec_obs_stab}. Owing to accretion, the star is
spinning up along the boundary of the stability peak
produced by the avoided crossing of modes I and II. This
stage is discussed in detail below. At point $D$ the star,
for the first time, becomes unstable with respect to
excitation of the octupole $m=3$
$r^o$-mode.%
\footnote{In principle, the magnetic field can limit
accretion spin-up before reaching point $D$
\cite{rfs04,gck14_short}.\label{footnote_mag} }
The amplitude of this mode increases rapidly and hits
saturation, which leads to heating up of the star. As a
result, it leaves the stability peak, becomes unstable also
with respect to excitation of mode I, and quickly moves to
point $E$. Thus, the $D$--$E$ stage is quite similar to the
$B$--$C$ stage of the standard scenario with the only
difference that the two modes ($m=3$ $r^o$-mode and the
mode I) are excited (and saturated) in this stage instead
of one. The spin frequency is almost constant during this
stage. At point $E$, the star approaches the
Cooling=Heating curve and then spins down along this curve
until it enters the stability region at point $F$ (stage
$E$--$F$). All the oscillation modes vanish in the very
beginning of the subsequent stage $F$--$A$ and the star
cools down to the equilibrium temperature $T_{\rm
eq}^\infty$ without noticeable variation of the spin
frequency. Stages $E$--$F$ and $F$--$A$ are close analogues
of, respectively, stages $C$--$D$ and $D$--$A$ of the
standard evolution scenario (see Fig.\
\ref{Fig_typical_obs}).

Let us return to the almost vertical stage $C$--$D$ in
Fig.\ \ref{Fig_scenario} and discuss it in more detail.
During this stage, the NS moves along the instability curve
for mode II. Only mode II is excited; the amplitudes of
other modes are all equal to zero. Since in stage $C$--$D$
the stellar temperature $T^\infty>T^\infty_\mathrm{eq}$,
the star requires an additional heating to maintain its
thermal balance. This heating is provided by the damping of
mode II. A required power is determined from Eq.\
(\ref{therm_i}) by setting $dT^\infty/dt \approx 0$,
\begin{equation}
  W_{{\rm Diss} \, \mathrm{II}} \approx L_{\rm cool} - K_{\rm n} \, \dot{M}\,
  c^2.
  \label{alpha1}
\end{equation}
Using Eqs.\ (\ref{Wdiss_i}) and (\ref{alpha1}),
together with Eqs.\ (\ref{Jc2}) and (\ref{Ec}),
one can determine the corresponding
{\it equilibrium} oscillation amplitude
\begin{equation}
 \alpha^{\rm (eq)}_{\mathrm{II}}
  \approx \sqrt{\frac{ (L_{\rm cool}-K_{\rm n}\, \dot{M}\, c^2) \, \tau_{{\rm Diss\, II}}}
  {\widetilde{J}\, M \, R^2\, \Omega^2}},
  \label{alpha2}
\end{equation}
where $\widetilde{J} \approx 1.6353 \times 10^{-2}$ for
mode II. Since $\tau_{{\rm Diss}\, {\rm II}}=-\tau_{{\rm
GR}\, {\rm II}}$ on the instability curve, one can use
$-\tau_{{\rm GR}\, {\rm II}}$
instead of $\tau_{{\rm Diss}\, {\rm II}}$ in this equation.%
\footnote{ It is convenient to use $-\tau_{{\rm GR}\, {\rm
II}}$ instead of $\tau_{{\rm Diss}\, {\rm II}}$ in Eq.\
(\ref{alpha2}), since $\tau_{{\rm Diss}\, {\rm II}}$ is a
strong function of $T^\infty$ in the vicinity of the
stability peak. The reason is the increasing role of the
mutual friction dissipation owing to an admixture of the
superfluid mode to the real solution near avoided crossing
(see Sec.\ \ref{Subsec_mix}). Thus, one cannot estimate
$\tau_{{\rm Diss}\, {\rm II}}$ directly from Eq.\
(\ref{tauS2}). On the opposite, the simple Eq.\
(\ref{tauGR2}) provides an accurate estimate for
$\tau_{{\rm GR}\, {\rm II}}$ because the gravitational
radiation time scale for the normal mode is smaller than
for the superfluid one. Hence, an admixture of the
superfluid mode has almost no effect on the gravitational
time scale for the real NS mode II [see Eq.\
(\ref{tau_resB})]. }
For example, taking the point on the instability curve with
coordinates $\nu=400$~Hz and $T^\infty \approx 1.48 \times
10^8$~K, we obtain $L_{\rm cool} \approx 3.3 \times
10^{34}$~erg\,s$^{-1}$; $\tau_{{\rm Diss}\, {\rm II}}=
-\tau_{{\rm GR}\, {\rm II}} \approx 1.13 \times 10^4$~s;
and, as follows from Eq.\ (\ref{alpha2}), $\alpha^{\rm
(eq)}_{\mathrm{II}} \approx 8 \times 10^{-7} \ll\alpha_{\rm
sat \, II}=10^{-4}$
(note that, when climbing the peak $L_\mathrm{cool}$ stays
almost constant; hence, the equilibrium amplitude scales as
$\alpha^{\rm (eq)}_{\mathrm{II}}\propto \nu^{-4}$, because
$\tau_{\rm GR \, II}\propto \nu^{-6}$).

It is possible for a star to maintain a finite, but not
saturated oscillation amplitude for a long time, because it
penetrates into the instability region with decreasing
$T^\infty$. Indeed, if, for some reason, mode II has a
lower amplitude than that required by Eq.\ (\ref{alpha2}),
then the star starts to cool down and becomes unstable with
respect to excitation of mode II. This immediately leads to
increasing of the amplitude $\alpha_{\rm II}$ and to
accelerated heating of the star. As a result, the star
moves toward the stability region, where $\alpha_{\rm II}$
decreases rapidly, the heating becomes less and less
efficient and, eventually, heating is replaced by cooling.
The process of modulation of $\alpha_{\rm II}$ may occur
repeatedly, but the correspondent variation of $T^\infty$
is very small. The characteristic modulation period varies
from a few months to years.

It can be shown that the modulation magnitude may decrease
or increase in time depending on the parameters of the
model. In the first case, during the NS motion along the
peak, the amplitude of mode II adjusts itself to the
equilibrium value $\alpha_{\rm II} \approx \alpha^{\rm
(eq)}_{\mathrm{II}}$ and does not experience modulation. In
the second case, the maximum value of $\alpha_{\rm II}$ is
typically limited by the saturation amplitude ($\alpha_{\rm
II}=\alpha_{\rm sat \, II}$), thus limiting the modulation
magnitude. However, even in this case the temperature
oscillations accompanying the modulation are very small,
less than the thickness of the line in Fig.\
\ref{Fig_scenario},
and can hardly be observed.%
\footnote{ The thermal relaxation of a NS crust can also
smooth the temperature oscillations.}
At the same time,
strong modulation of the oscillation amplitude $\alpha_{\rm II}$
is also accompanied by the modulation of $d \Omega/dt$,
which is, in principle, observable.%
\footnote{Note that only the period of modulation and its
magnitude depend on the shape of the instability curve; in
contrast, the fact that the star stays attached to this
curve is purely due to the onset of gravitational
instability with decrease of $T^\infty$. Consequently, the
exact form of the instability curve and the function
$\theta(x)$, which determines it [see Eq.\ (\ref{theta})],
are insignificant for our model.}
The effects of $\alpha_{\rm II}$ modulation described above
will be discussed in detail in our subsequent publication.

Let us estimate the duration of the spin-up stage $C$--$D$.
Using Eqs.\ (\ref{Omega_i}) and (\ref{alpha2}), we get
\begin{equation}
\frac{d \Omega}{dt} = - \frac{2 \, Q_{\rm II} \, (L_{\rm
cool}-K_{\rm n}\, \dot{M}\, c^2)} {\widetilde{J} \, M \, R^2 \,
\Omega} +\dot{\Omega}_{\rm acc}, \label{Omega2}
\end{equation}
where $Q_{\rm II}\approx 0.094$ [see Eq.\ (\ref{Q})].
First of all,
taking into account Eq.\ (\ref{Omega_acc})
one can determine from this formula
the minimal NS
accretion rate $\dot{M}_{\rm min}$
required to spin up
the star,
\begin{eqnarray}
\dot{M}_{\rm min} &=& \frac{3 L_{\rm cool}}{3 \, K_{\rm n} \, c^2 +p
\, \Omega\,  \sqrt{G\, M\, R}}
\nonumber\\
&\approx& \frac{3 \times 10^{-12}}{p} \,
\left(\frac{L_{\rm cool}}{10^{34} \, {\rm erg \, s}^{-1}}\right) \,
\left(\frac{\Omega}{\Omega_0}\right)^{-1} \,
\left(\frac{M}{1.4 \, M_\odot}\right)^{-1}  \left(\frac{R}{10 \, {\rm km}} \right) \,
\frac{M_{\odot}}{{\rm yr}}.
\label{Mdotmin}
\end{eqnarray}
At point $C$, one has $\Omega_C \approx 1500$~s$^{-1}$
($\nu_{C} \approx 239$~Hz), $T_C^\infty \approx 1.37 \times
10^{8}$~K, $L_{\rm cool} \approx 3 \times 10^{34}$~erg
s$^{-1}$, and it follows from Eq.\ (\ref{Mdotmin}) that
$\dot{M}_{\rm min}\approx 6.2 \times 10^{-11}
M_{\odot}/{\rm yr}$. If $\dot{M}>{\dot M}_{\rm min}$, then
the duration of the $C$--$D$ stage can be estimated by
noticing that the first term in Eq.\ (\ref{Omega2}) is
smaller than the second one at $\Omega \ga \Omega_C$.
Because $\Omega_D \approx 3930$~s$^{-1}$ ($\nu_D \approx
625$~Hz), we find
\begin{equation}
\Delta t_{CD} \approx
\frac{\Omega_D-\Omega_C}{\dot{\Omega}_{\rm acc}} \approx
2.2 \times 10^8 \, {\rm yr}, \label{tau_vverh}
\end{equation}
where we make use of Eq.\ (\ref{Omega_acc}) with our
fiducial accretion rate $\dot{M}=3\times 10^{-10}
M_\odot/{\rm yr}$. An accurate calculation, which is done
without any additional simplifications, gives a close value
$\Delta t_{CD} \approx 2.3\times 10^8$~yr. This time
constitutes approximately $82\%$ of the period of the
$A$--$B$--$C$--$D$--$E$--$F$--$A$ cycle. For comparison,
the $A$--$B$ and $E$--$F$ stages constitute, respectively,
$15\%$ and $3\%$ of the cycle; the contribution of all
other stages is negligible. Note that the time $\Delta
t_{CD}$ can be even longer, if the magnetodipole torque is
sufficiently large. The corresponding term of the form
\begin{equation}
\dot{\Omega}_{B}=-\frac{B^2 \, R^6}{6 \, c^3 \, I} \,
\Omega^3
\label{B}
\end{equation}
should then be added to the right-hand side of Eq.\
(\ref{Omega2}). In particular, for a strong enough dipolar
magnetic field $B$, a star can stop spinning up at a
frequency at which $\dot{\Omega}_{\rm acc}+\dot{\Omega}_{B}
\approx 0$. For example, this will happen at $\nu=600$~Hz
[for $\dot{M}=3\times 10^{-10} M_\odot/{\rm yr}$ and
accretion torque given by Eq.\ (\ref{Omega_acc})] if the
magnetic field at the poles is $B \approx 8.8 \times
10^8$~G.

Four conclusions can be drawn
from the analysis of Fig.\ \ref{Fig_scenario}
and the estimates presented above.

($i$) The high spin frequencies of the sources 4U 1608-522,
SAX J1750.8-2900, EXO 0748-676, Aql X-1, and SWIFT
J1749.4-2807 can be explained assuming that these stars are
climbing up the peak in the $C$--$D$ stage;

($ii$) The probability to find these stars with the
observed (high) frequencies is not small,
since they spend a
substantial amount of time in the high frequency region;

($iii$) The maximum NS spin frequency is limited by the
$m=3$ $r^o$-mode instability curve within our scenario;

($iv$) A star, which starts to evolve in the stability
region with the temperature lower than that of the avoided
crossing of modes I and II, will eventually find itself in
stage $C$--$D$.

\begin{figure}
    \begin{center}
        \leavevmode
        \includegraphics[width=3.4in]{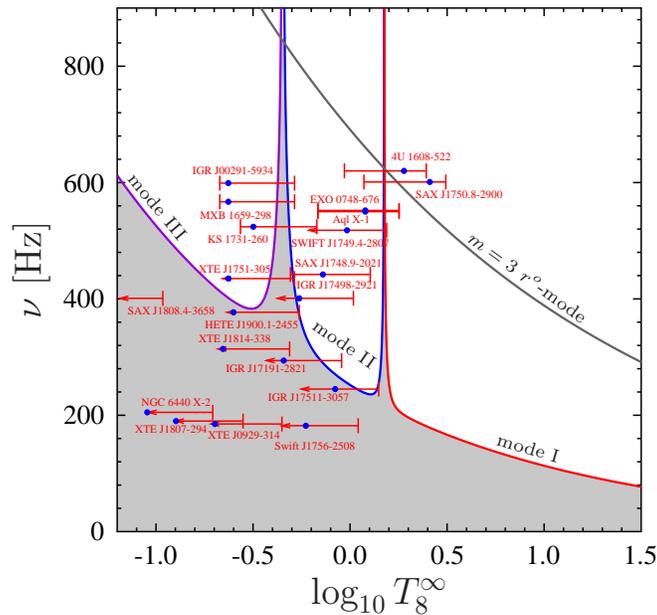}
    \end{center}
    \caption{(color online) An example of the stability curves in case
    of two avoided crossings of $m=2$ oscillation modes of
    a superfluid NS.
    As in Fig. \ref{Fig_instab},
    the solid  lines
    are plotted for  modes I and II (red and blue online, respectively) experiencing an avoided crossing
    at $T^\infty=1.5\times10^8$~K (the coupling parameter $s=0.001$).
    An additional solid line
    (violet online) corresponds to  mode III,
    which exhibits an avoided crossing with the mode II
    at $T^\infty=4.5\times10^7$~K.
    This avoided crossing is drawn for $s=0.01$.
    Other notations are the same as in Fig.\ \ref{Fig_instab}.
    }
    \label{Fig_2peak}
\end{figure}

The other sources with lower $T^\infty$ (e.g., IGR
J00291-5934) can be explained in a similar manner. First,
it is obvious that the temperature $T^{\infty}_0$ of the
avoided crossing of modes I and II depends on the NS mass.
Hence, if the masses of colder sources differ from those of
the hotter ones, the avoided crossing of modes I and II can
occur at a different $T^{\infty}_0$. In particular, it can
be shifted to the region of lower temperatures, which are
typical for these (rather cold) stars. Second, as it was
shown in calculations of nonrotating NS oscillation spectra
\cite{ga06, kg11, cg11, gkcg13,gkgc14}, a normal mode can
experience an avoided crossing with the superfluid modes
more than once. To illustrate this idea, we demonstrate in
Fig.\ \ref{Fig_2peak} the instability curves in the case of
two avoided crossings of oscillation modes. The first
avoided crossing takes place at $T^\infty=4.5 \times
10^7$~K between mode III
(solid line marked ``mode III'' in the figure; violet
online), which behaves as an $m=2$ $r^o$-mode at low
$T^\infty$, and mode II
(solid line; blue online).
For this avoided crossing the coupling parameter was chosen
to be $s=0.01$. The second avoided crossing of modes I and
II is discussed above (see Fig.\ \ref{Fig_scenario}); it
takes place at $T^\infty=1.5\times10^8$~K. In this case
mode II behaves as $m=2$ $r^o$-mode only at intermediate
temperatures $6\times 10^7\ \mathrm{K}\lesssim T^\infty
\lesssim 1.3\times 10^8$~K. At higher and at lower
temperatures it transforms into superfluid modes, which
are, generally, different.
It is easy to demonstrate that, for low enough
$T_\mathrm{eq}^\infty\lesssim 4\times 10^7$~K, the
evolution track goes along the left (low-temperature)
boundary of the first stability peak, corresponding to the
avoided crossing of modes II and III [i.e., along
the ``mode III''
line (violet online) in Fig.\ \ref{Fig_2peak}]. This stage
is a direct analogue of the $C$--$D$ stage in Fig.\
\ref{Fig_scenario}, and a NS stays there for a long time.
One sees that two avoided crossings%
\footnote{In reality, the
number of avoided crossings can be larger.}
are already sufficient
to explain all the existing
observations of frequencies
and quiescent temperatures
of NSs in LMXBs.

In summary,  the sources IGR J00291-5934, MXB 1659-298, KS
1731-260, and XTE J1751-305 can be interpreted as moving
along the instability curve of mode III (the curve which is
violet online in the figure). This interpretation requires
that their equilibrium temperature
$T_\mathrm{eq}^\infty\lesssim 4\times 10^7$~K. The
parameters of the objects IGR J17498-2921 and SAX
J1748.9-2021 can be explained by accretion spin-up at
$T^\infty=T_\mathrm{eq}^\infty\sim 5\times 10^7$~K, which
takes place inside the stability region (an analogue of the
$A$--$B$ stage in the scenarios discussed above). Finally,
an explanation of the hottest stars 4U 1608-522, SAX
J1750.8-2900, EXO 0748-676, Aql X-1, and SWIFT J1749.4-2807
remains the same as in the scenario with one avoided
crossing (however, because of the additional avoided
crossing the equilibrium temperature $T_\mathrm{eq}^\infty$
should comply with the condition $5\times 10^7\
\mathrm{K}\lesssim T_\mathrm{eq}^\infty \lesssim 1.4\times
10^8$~K for these sources). The rest of the stars lie in
the stability region (even without accounting for the
resonant coupling of modes), so they can be explained as
being in the $A$--$B$ stage with the corresponding
equilibrium temperature $T^\infty_{\rm eq}$ (see Sec.\
\ref{Subsec_obs_stab}).

Let us note that, to spin up the rapidly rotating sources
up to the observed spin frequencies $\Omega$ during the
time period shorter than the age of the Universe
$t_\mathrm{Un}$, one needs quite a strong accretion torque
$\dot \Omega_\mathrm{acc}^{\rm (crit)}\gtrsim
\Omega/t_\mathrm{Un}\sim 3\times
10^{-7}$~s$^{-1}$\,yr$^{-1}$, and hence quite a high
accretion rate [$\dot{M}_{\rm crit}\gtrsim 10^{-11}
M_\odot/\mathrm{yr}$ if one uses Eq.\ (\ref{Omega_acc})].
Thus, to explain the low-temperature sources (those like
MXB 1659-298) a rapid NS cooling may be required (e.g.,
with the open direct Urca process in the central regions of
the star; see Ref.\ \cite{ykgh01} and Sec.\ \ref{MSP}),
which allows one to have a lower $T_\mathrm{eq}^\infty$ at
higher ${\dot M}$. This can indicate that the coldest
rapidly rotating NSs in LMXBs are more massive. An
alternative explanation of these sources (not requiring an
enhanced cooling) is also possible. It implies a more
efficient accretion torque
for these objects,
that results in a large value of $\dot \Omega_\mathrm{acc}$
at a relatively small accretion
rate~${\dot M}$.
The last hypothesis
agrees
with the
very low observational estimate $\dot{M} \approx 2.5 \times
10^{-12} \, M_\odot/{\rm yr} \ll \dot{M}_{\rm crit}$ for
the source IGR J00291-5934
(see Table \ref{Tab_LMXB_Observ}),%
\footnote{Average accretion rate estimated by Patruno
\cite{patruno10} is three times larger $\dot{M} \sim
(7$--$8) \times 10^{-12} \, M_\odot/{\rm yr}\sim
\dot{M}_{\rm crit}$.}
as well as with the results of Ref.\ \cite{ajh14}, in which
it is shown that the high spin-up rates observed for XTE
J1751-305 and IGR J00291+5934 are not quite consistent with
theoretical estimates.
In Fig.\ \ref{Fig_2peak} we have considered a situation in
which an additional avoided crossing appears at lower
$T^\infty$ than for modes I and II. It is also interesting
to see how the additional avoided crossing affects the NS
evolution if it appears at higher $T^\infty$. This
possibility is studied in Appendix \ref{Appendix_2peaks},
where it is shown that the four conclusions ($i$)--($iv$)
stated above hold true even in this case.

\section{
Neutron star evolution after the end of accretion phase
and production of millisecond pulsars
\label{MSP}}

Thus, we demonstrate that the high spin frequencies of NSs
in LMXBs can naturally be explained within our new
scenario. But is this scenario compatible with the
existence of millisecond pulsars (MSPs)? It is generally
believed \cite{acrs82} that MSPs originate from LMXBs, in
which accretion has ceased for some reason, for example,
because of a binary system evolution \cite{Tauris11,tvh06}
or close encounter with some other star
\cite{ivanova_etal08}. Let us consider the evolution of a
NS with accretion switched off. A few alternatives are
possible.

\begin{itemize}

\item Accretion ceases when a NS is in stage $A$--$B$
or in a similar stage with lower $T^\infty_{\rm eq}$.
%
\footnote{NSs in stages $D$--$E$, $E$--$F$, and $F$--$A$
(or in their analogues associated with the low-temperature
stability peak)
will eventually find themselves in the stability region
with the frequency $\Omega=\Omega_A$, independently of
whether they are accreting or not (see Appendix
\ref{Appendix_StandScen} for the definition of $\Omega_A$).
Their subsequent evolution is then similar to what is
discussed here.}

Then the star is stable and CFS instability does not affect
its evolution. As a result, the NS cools down rapidly,
keeping its spin frequency almost unchanged and eventually
becomes a MSP. In this formation channel NS spin
frequencies are limited by the instability curve. For
realistic $T^\infty_\mathrm{eq}\ga 4 \times 10^7$~K
this means that only MSPs with spin frequencies up to
$\nu\lesssim 400$~Hz can be formed in this way (see Fig.\
\ref{Fig_2peak}). To form even faster MSPs, with $\nu$ up
to $500$~Hz, one should assume that they had lower
equilibrium temperatures ($T_{\rm eq}^\infty \sim 10^7$~K)
in the $A$--$B$ stage. This is possible (see below),
provided that these stars are massive enough so that strong
neutrino emission processes (such as nucleon and/or hyperon
direct Urca processes) are opened in their cores.

\item Accretion ceases when a NS is climbing up the high-temperature peak
in the $C$--$D$ stage (see Fig.\ \ref{Fig_scenario}).

%
In that case a NS remains attached to the stability peak
because its cooling would make the CFS instability stronger
and heat the star up (the same situation as in the LMXB
system with accretion; see Sec.\ \ref{Sec_climb}). When
accretion ceases, the equilibrium amplitude $\alpha^{({\rm
eq})}_{\rm II}$ for mode II increases slightly [see Eq.\
(\ref{alpha2}) with $\dot M=0$].
Since in the absence of accretion $\dot{\Omega}_{\rm
acc}=0$, NS spin frequency will gradually decrease as the
rotation energy is carried away by gravitational waves and
neutrinos. To get an impression about the typical times of
climbing down, let us estimate the time $\Delta t_{DC}$
spent by a NS on the way from point $D$ to point $C$. Even
for a very high $T^\infty=1.5 \times 10^8$~K (and hence
$L_{\rm cool} \approx 3.45 \times 10^{34}$~erg s$^{-1}$),
Eq.\ (\ref{Omega2}) with $\dot{\Omega}_{\rm acc}=0$ and
$\dot{M}=0$ gives
\begin{equation}
\Delta t_{DC} = \frac{\widetilde{J} \, M \, R^2 \,
(\Omega_D^2-\Omega_C^2)} {4 \, Q_{\rm II} \, L_{\rm cool} }
\approx 1.5 \times 10^9 \, {\rm yr}. \label{DC}
\end{equation}
Such a long time indicates that the probability to observe
a rapidly rotating nonaccreting NS climbing down the peak
is not small. It is easy to demonstrate that, due to
magnetodipole losses only [see Eq.\ (\ref{B})], a star
would spin down during the same period of time if it had
the dipolar magnetic field at the poles $B \approx 7 \times
10^8$~G.

In Refs.\ \cite{gck14_short,cgk14} it is argued that these
nonaccreting NSs, heated by the CFS instability, form a
specific new class of NSs. These references propose to call
them ``HOFNARs'' (from HOt and Fast Non-Accreting Rotators)
or ``hot widows'' (in analogy with the ``black widow''
pulsars), and suggest that a number of sources that are
tentatively identified as quiescent LMXB candidates may in
fact be such objects.

Could ``hot widows''/HOFNARs, descending the
high-temperature peak, be associated with MSPs?
Most probably not, because these objects are very hot, with
effective surface temperature $T_{\rm eff}^\infty \sim
10^6$~K (and internal temperature $T^\infty \sim 10^8$~K),
while it is customary to assume that MSPs are much colder
(only their hot spots can reach the values $\sim 10^6$~K).
The high temperature of ``hot widows''/HOFNARs explains,
most likely, the fact that these objects do not show radio
pulsar activity: The magnetic field in hot NSs decays much
faster (see, e.g., Refs.\ \cite{uk08,cgk14}).
The detailed analysis of this possible new class of NSs
from both theoretical and observational points of view is
presented in Ref.\ \cite{cgk14}.

\item Accretion ceases when a NS is climbing up the low-temperature peak
(like the left peak in Fig.\ \ref{Fig_2peak}
or a similar peak at lower temperature).

The subsequent evolution of a NS is then quite similar to
that in the case of the high-temperature peak. The star
becomes a ``hot widow''/HOFNAR; the only difference is that
now its temperature is noticeably smaller. As a
consequence, such star can maintain its magnetic field and
be, at the same time,
a MSP. Therefore, MSPs with spin frequencies $\nu \gtrsim
(400$--$500)$~Hz (including the most rapidly rotating
pulsar PSR J1748-2446ad with $\nu = 716$~Hz) are
interpreted by us as NSs, climbing down the low-temperature
stability peak.

Here it is pertinent to ask the following question: What is
the minimal possible temperature $T_0^\infty$ of the
stability peak, at which a NS still can find itself there?
Obviously, for that to be possible, the equilibrium
internal temperature $T_{\rm eq}^\infty$ of the star should
be smaller than $T_0^\infty$. This temperature is found
from the condition $L_{\rm cool}=K_{\rm n} \dot{M} c^2$. It
can be shown that even for $\dot{M}=\dot{M}_{\rm crit}$ and
completely unsuppressed nucleon direct Urca process (which
is quite unrealistic) $T_{\rm eq}^\infty \ga 6 \times
10^6$~K, which corresponds to the effective surface
temperature $T_{\rm eff}^\infty \ga 3 \times 10^5$~K (for
$P/g=10^9$~g cm$^{-2}$, see Sec.\ \ref{Subsec_obs}). In a
more realistic case, when we have a completely unsuppressed
direct Urca process with $\Lambda$ hyperons ($\Lambda
\rightarrow p+e+ \bar{\nu}_{e}$; see, e.g., Ref.\
\cite{ykgh01}), operating in the inner half of the NS core
($r \leq R/2$),
\footnote{ The suppression of this process by superfluidity
should be weak in the central NS regions, because proton
superconductivity is reduced considerably at large
densities \cite{plps09}, while recent microscopic
calculations predict that the critical temperatures for
$\Lambda$-hyperons are likely to be very small
\cite{tmc03,tnyt06,ws10}. }
one obtains
$T_{\rm eq}^\infty \sim 1.3 \times 10^7$~K,
which corresponds to
$T_{\rm eff}^\infty \sim 4.6 \times 10^5$~K.

One sees that both estimates give rather large values of
minimal equilibrium temperature. As a consequence, the
effective surface temperatures of MSPs with the spin
frequencies $\nu \gtrsim 400$--$500$~Hz, which, according
to our scenario, are climbing down the low-temperature
stability peak, cannot be lower than
$T_{\rm eff}^\infty \sim (3$--$5) \times 10^5$~K.%
\footnote{A hypothesis that rapidly rotating MSPs are
probably not so cold as it is generally believed agrees
with the observations of PSR J1723-2837 ($\nu\approx
539$~Hz; see Ref.\ \cite{Faulkner_etal04}) , which has
surface temperature $T^\infty_\mathrm{eff}\sim
(4$--$5)\times 10^5$~K \cite{bogdanov_etal14}.
\label{foot_HotMSP}}
It is important to note that this conclusion can change in
a more complicated scenario which accounts for a possible
resonance interaction of the core $r$-modes with elastic
modes of the crust \cite{lu01,yl01,ga06a}. In such scenario
a NS can leave, under certain conditions, the stability
peak and cool down to very low temperatures (see Appendix
\ref{Sec_MSP_Levin}). Note also that accounting for the
interaction of $r$-modes with the crust modes allows one to
explain cold MSPs with $\nu\gtrsim (400$--$500)$~Hz without
invoking powerful neutrino emission processes (such as the
direct Urca process) in the NS core.

\end{itemize}

In conclusion, the proposed scenario can explain the
formation of MSPs, including the most rapidly rotating
pulsars. It also predicts the existence of a new class of
hot and rapidly rotating NSs -- ``hot widows''/HOFNARs (see
Ref.\ \cite{cgk14} for details).
%

\section{Conclusions}
\label{Sec_concl}

We demonstrate that the key role in the evolution of NSs in
LMXBs is played by the resonance interaction of the normal
$m=2$ oscillation $r$-mode ($r^o$-mode) and the superfluid
inertial modes ($i^s$-modes).
This result allows us to
formulate a scenario that explains observations of rapidly
rotating warm NSs in LMXBs (Sec.\ \ref{Sec_climb})
and predicts the existence of a new class of
nonaccreting NSs which we propose to call ``hot widows'' or
HOFNARs (see Sec.\ \ref{MSP} and Ref.\ \cite{cgk14} for
more details).
This scenario is in agreement with the existence of MSPs
(Sec.\ \ref{MSP} and Appendix \ref{Sec_MSP_Levin}),
predicting that some of them [especially, most rapidly
rotating MSPs with $\nu \gtrsim (400$--$500)$~Hz] can be
rather hot, with the effective surface temperatures $T_{\rm
eff}^\infty \ga (3$--$5) \times 10^5$~K.
A more detailed analysis of our scenario in application to
MSPs will be reported elsewhere.

The conclusion about
the resonance interaction
of $i^s$- and $r^o$-modes is based on the following facts:

$1.$ Detailed calculations \cite{ga06,gk11, kg11, cg11,
gkcg13,gkgc14} of the oscillation spectra of nonrotating
superfluid NSs at finite temperatures $T^\infty$ reveal
that ($i$) The frequencies of the superfluid modes
essentially depend on $T^\infty$, while those of the normal
modes are almost insensitive to a temperature variation.
($ii$) If, at some $T^\infty$, the frequencies $\omega$ of
two arbitrary (but with the same ``quantum'' number $m$)
superfluid and normal modes become equal, they start to
interact resonantly. As a result of such interaction, the
superfluid mode turns into the normal one and vice versa;
that is, an avoided crossing of modes is formed in the
$\omega-T^{\infty}$ plane. ($iii$) Far from the avoided
crossings superfluid and normal modes are almost
noninteracting and are described by the two weakly coupled
systems of equations.

$2.$ According to computations
of Lee and Yoshida \cite{ly03,yl03a}
performed in the $T^\infty=0$ approximation,
the frequencies of $i^s$-modes are sensitive
to a variation of the so-called
entrainment parameter $\widetilde{\eta}$
(see Sec.\ \ref{Subsec_2assum}).
In particular, at some specific values of $\widetilde{\eta}$
avoided crossings of superfluid and normal
oscillation modes are observed
(see also Sec.\ \ref{Subsec_mix}).

$3.$ An account for finite $T^\infty$
leads to a temperature dependence
of a number of parameters
of superfluid hydrodynamics (including $\widetilde{\eta}$).

Items $2$ and $3$ give us a ground to {\it assume} that the
results formulated in item $1$ in application to
nonrotating NSs remain valid for rotating NSs as well.
Hence, the frequencies of $i^s$-modes should also depend on
$T^\infty$. This means, in particular, that avoided
crossings between the $m=2$ $r^o$-mode and $i^s$-modes
should be formed at some values of $T^\infty$ (see, e.g.,
Fig.\ \ref{Fig_scheme}). When passing through an avoided
crossing, the $i^s$-mode transforms into the $m=2$
$r^o$-mode, while the $m=2$ $r^o$-mode becomes the
$i^s$-mode. During such a transformation the eigenfunctions
of the $m=2$ $r^o$-mode mix intensively with those of the
$i^{s}$-mode. This leads to the enhancement of $r^o$-mode
damping due to mutual friction (see Sec.\
\ref{Subsec_mix}). In the $\nu-T^\infty$ plane, this effect
is manifested by the appearance of a sharp ``stability
peak'' over the standard (usually considered) stability
region of fast rotating NSs (see Sec.\ \ref{Subsec_wind}
and Fig.\ \ref{Fig_instab}; the stability region is filled
with grey there).

An analysis of evolution of a NS in LMXB taking into
account the stability peak shows that the star spends a
significant amount of time climbing the left side of this
peak in the region, which has been previously thought to be
unstable with respect to excitation of $r$-modes. To keep
on the peak, the average oscillation amplitude adjusts
itself so that the star heating due to dissipation of
oscillations is compensated by neutrino cooling. Under such
circumstances, a spin-down due to gravitational wave
emission can be insufficient to oppose the accretion torque
on the star. This leads to a gradual increasing of the NS
spin frequency as it slowly climbs up the peak (see Sec.\
\ref{Sec_climb} for details).

If spin-up is not terminated by the magnetic field (see
footnote \ref{footnote_mag}),
the star reaches the instability curve for the $m=3$
oscillation $r^o$-mode, which is the next unstable mode in
normal NSs
after the $m=2$ $r^o$-mode.%
\footnote{The possibility that, under certain
circumstances, another (secular or dynamical) instability
could set in at lower $\Omega$ than the instability of the
octupole ($m=3$) $r^o$-mode cannot be excluded and should
be carefully analyzed.}
As a result, the star jumps off the peak and shortly
returns to the stability region (see Sec.\
\ref{Sec_climb}). Thus, the real limit on the spin
frequency of NSs is set by the instability curve for the
octupole $m=3$ $r^o$-mode. This result allows us to explain
the fast rotation of NSs in LMXBs within the minimal
assumptions about the properties of superdense matter.
Moreover, this result agrees with the predicted
\cite{chakrabarty_etal_03, chakrabarty08} abrupt cutoff
above $\sim 730$~Hz of the spin frequency distribution of
accreting millisecond X-ray pulsars. Furthermore, because
$\dot \Omega\approx \dot \Omega_{\rm acc}$ in the $A$--$B$
and $C$--$D$ stages, our scenario predicts the frequency
distribution to be almost constant at $200$--$600$~Hz, in
agreement with observations (see, e.g., Fig.\ 5 of Ref.\
\cite{patruno10}).

It is important to emphasize that our scenario is almost
insensitive to an actual choice of the parameters
regulating the resonance interaction between the modes
(resonance temperatures and width of the peaks; see Secs.\
\ref{Sec_sflnorm} and \ref{Sec_wind_regime}) and does not
require any nonrealistic enhancement of the kinetic
coefficients and/or additional exotic damping mechanisms.

Obviously, the new approach to the evolution of rapidly
rotating NSs and interpretation of their observations,
suggested in the present paper, needs further development
and refinement. In particular, one needs to perform
detailed calculations in order to confirm the presence of
avoided crossings in the oscillation spectra of warm
superfluid rotating NSs, and to study how the resonance
interaction of modes affects the oscillation damping times.
We expect that the corresponding resonance temperatures
will depend on the NS mass and on the parameters of
superfluidity. A detailed analysis of the effect of various
damping processes (such as, e.g., Ekman layer dissipation
\cite{bu00,Rieutord01,Rieutord01_erratum,ak01,ga06b}) on
the instability curve of the octupole $m=3$ $r^o$-mode will
place further restrictions on the spin frequencies of NSs.

If our scenario is correct, then the observed temperatures
of the most rapidly rotating NSs must coincide with the
temperatures $T^\infty_0$, at which avoided crossings occur
between the $m=2$ $r^o$-mode and the superfluid
$i^s$-modes. Comparison of these temperatures $T^\infty_0$
with the results of (still not available) theoretical
calculations can impose stringent constraints on the
properties of superdense matter and parameters of
superfluidity. Clearly, a direct observational test of our
scenario is a very important task which we plan to address
in the nearest future. In particular, we plan to study in
detail the modulation of the NS spin frequency, appearing
when the star moves along the stability peak (see Sec.\
\ref{Sec_climb}), and to examine whether this effect can be
confirmed observationally.

\section*{Acknowledgements}

 We are grateful to
A.~A. Danilenko, A.~D. Kaminker, O.~Y. Kargaltsev, G.~G.
Pavlov, A.~Y. Potekhin, Y.~A.~Shibanov, A.~I. Tsygan,
V.~A.~Urpin, D.~G. Yakovlev, and D.~A. Zyuzin for
insightful comments and discussions, and to O.~V.
Zakutnyaya for assistance in preparation of the manuscript.
This work was partially supported by RFBR (grants
14-02-00868-a and 14-02-31616-mol-a), by RF president
programme (grants MK-506.2014.2 and NSh-294.2014.2), and by
the Dynasty Foundation.

\appendix

\section{The approximations for neutrino luminosity and heat capacity}
\label{Appendix_LC}

The neutrino luminosity $L_{\rm cool}$ and total heat
capacity  $C_{\rm tot}$ of a NS with the mass $M=1.4
M_{\odot}$ are calculated with the relativistic cooling
code, described in detail in Refs.\
\cite{gkyg04,gkyg05,yp04}. We use essentially the same
microphysics input as in Ref.\ \cite{gkyg04}. In
particular, we employ the parametrization \cite{hh99} of
APR EOS \cite{apr98}. The results of our calculations of
$L_{\rm cool}$ and $C_{\rm tot}$ are roughly fitted by the
following formulas
\begin{eqnarray}
L_{\rm cool}&=&7\times 10^{30}\left(T^\infty_8\right)^8
\,\left\{\sqrt{1.25\,T^\infty_8}
        +140\,\exp\left(-\frac{30}{T_8^\infty}\right)\right.\nonumber \\
        &+&\left.3\times 10^4 \,\exp\left[-\left(\log_{10}T_8^\infty+0.5\right)^2/0.3^2\right]
        \right\}\, \mathrm{erg\, s}^{-1},
        \\
C_{\rm tot}&=&2.1\times10^{37}\,T^\infty_8\,\left(1+\frac{6}{(0.18/T^\infty_8)^{3.6}+1}\right) \,
{\rm erg} \, {\rm K}^{-1}.
\label{C_Lcool_approximation}
\end{eqnarray}
The last term in the expression for $L_{\rm cool}$
corresponds to the enhancement of the neutrino luminosity due to
neutron Cooper pairing.

\section{NS evolution in the absence of resonant
interaction with superfluid modes
}\label{Appendix_StandScen}

\begin{figure}
    \begin{center}
        \leavevmode
        \includegraphics[width=3.4in]{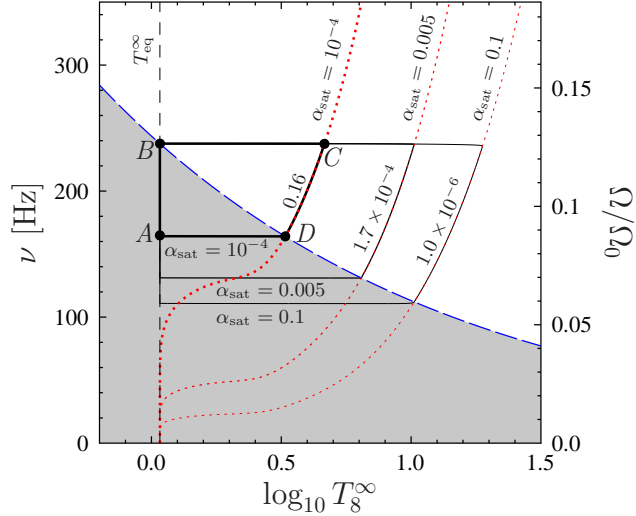}
    \end{center}
    \caption{(color online)
 Analogous to Fig. \ref{Fig_typical_obs},
 but in a larger scale.
 The evolution of
    the spin frequency $\nu$ and
    temperature $T^\infty_8$ is shown
    for a NS in LMXB
    in the absence of resonant interaction of modes.
    The corresponding tracks are shown by solid lines
    (black online; thick, medium, and thin lines
    are for $\alpha_\mathrm{sat}=10^{-4}$, $0.005$, and $0.1$, respectively).
    The numbers near the lines indicate the fraction of time
    the star spends in the instability region
    (this fraction is calculated without accounting for the time
    $\Delta t_{\rm torq}\approx 4500$~yr,
    during which the star is located at point $B$).
    The stability region of the $m=2$ $r^o$-mode is filled with
    grey;
    its boundary is shown by a thick dashed line (blue online).
    Vertical dashed line shows the
    equilibrium stellar temperature $T_\mathrm{eq}^\infty$.
    The dotted curves are plotted assuming that the neutrino cooling
    exactly balances the stellar heating due to nonlinear dissipation
    of the saturated $m=2$ $r^o$-mode (the Cooling=Heating curves).
    See text for details.
        }
    \label{Fig_typical}
\end{figure}
%

Let us analyze the evolution of a NS in LMXB in the absence
of resonant interaction of the normal $r^o$-mode with
superfluid modes. A similar scenario was proposed, for the
first time, in Ref.\ \cite{levin99} (see also Ref.\
\cite{heyl02}). Here we reconsider it, employing the
physics input, described in Sec.\ \ref{Sec_input}, and
perform a number of useful estimates, supplementing the
consideration of Sec.\ \ref{Subsec_obs_stab}. Figure
\ref{Fig_typical} presents the stellar spin frequency $\nu$
as a function of the internal redshifted temperature
$T^\infty$. The thick solid line shows the cyclic evolution
track of the NS $A$--$B$--$C$--$D$--$A$ for the saturation
amplitude of the $r^o$-mode $\alpha_{\rm sat}=10^{-4}$. The
medium and thin solid lines show similar tracks for
$\alpha_{\rm sat}=5 \times 10^{-3}$ and $\alpha_{\rm
sat}=10^{-1}$, respectively.

The instability curve for the quadrupole $m=2$ $r^o$-mode,
given by the condition $1/\tau_{\rm GR}+1/\tau_{\rm
Diss}=0$, is shown by a thick dashed line (blue online). In
the region over the curve one has $1/\tau_{\rm
GR}+1/\tau_{\rm Diss}<0$ and, as follows from Eq.\
(\ref{dadt}), the star is unstable with respect to
excitation of the $r^o$-mode ($d\alpha/dt>0$). In the
figure the stability region for the $m=2$ $r^o$-mode is
filled with grey.

Let us discuss in more detail the stellar evolution along
the track $A$--$B$--$C$--$D$--$A$.

($i$) Stage $A$--$B$.

The star has initial equilibrium temperature
$T_A^\infty=T_{\rm eq}^\infty \approx 1.078 \times 10^8$~K
(see Sec.\ \ref{Subsec_obs_stab}), the amplitude of the
$r^o$-mode $\alpha=0$, and the spin frequency $\nu_A
\approx 164$~Hz ($\Omega_A= 2 \pi \nu_A\approx
1030$~s$^{-1}$). The stellar spin frequency grows linearly
due to accretion of matter onto the NS according to Eq.\
(\ref{dOmegadt}) with $\alpha=0$. The stellar temperature
$T^\infty$ remains unchanged. The star reaches the boundary
of the stability region at point $B$. In this point $\nu_B
\approx 238$~Hz ($\Omega_B\approx 1495$~s$^{-1}$), so that
the time spent by the star in stage $A$--$B$ equals $\Delta
t_{AB}=(\Omega_B -\Omega_A)/\dot{\Omega}_{\rm acc} \approx
4 \times 10^7$~yr.

($ii$) The stage $B$--$C$.

At point $B$, the star is located on the instability curve
of the $m=2$ $r^o$-mode. Further increasing of the stellar
spin frequency makes it unstable. However, if the
$r^o$-mode amplitude is strictly zero, then, as follows
from Eq.\ (\ref{dadt}), $d \alpha/dt$ remains to be zero
even in the instability region. In reality, of course, any
fluctuation of the amplitude $\alpha$ (for instance, the
thermal fluctuation or a fluctuation related to accretion
onto a NS) will lead to instability growth. In numerical
calculations we modeled this effect by specifying the
initial condition $\alpha_{B}=10^{-30}$ for the oscillation
amplitude at point $B$. Naturally, the subsequent NS
evolution is not sensitive to the actual value of the
initial amplitude.

Becoming unstable, the amplitude of $m=2$ $r^o$-mode grows
rapidly, so that after $\Delta t_{\rm torq} \approx
4500$~yr the torque acting on the NS due to $r^o$-mode
dissipation becomes equal to the accretion torque [$d
\Omega/dt =0$; see Eq.\ (\ref{dOmegadt})]. This happens at
$\alpha_{0} \approx 1.8 \times 10^{-5}$. Approximately
$4$~yr later the $r^o$-mode reaches saturation
($\alpha=\alpha_{\rm sat}=10^{-4}$). During these evolution
phases $\Omega$ and $T^\infty$ almost do not change. The
spin frequency $\Omega$ does not change because a typical
time scale of its variation is much greater than $\Delta
t_{\rm torq}$ (see below). The temperature $T^\infty$ does
not change, because its typical time scale is $\propto
\alpha^{-2}$ [see Eqs.\ (\ref{thermal}) and (\ref{Wdiss})]
and is also much greater than $\Delta t_{\rm torq}$ most of
the time.

Using the fact that $\Omega$ and $T^\infty$ are almost
constant, $\alpha_0$ can be derived from Eq.\
(\ref{dOmegadt}) if we fix $\Omega=\Omega_B$ and
$T^\infty=T^\infty_{\rm eq}$ and make its left-hand side
vanish,
\begin{equation}
  \alpha_0=\sqrt{\frac{\dot{\Omega}_{\rm acc}
    \, \tau_{\rm Diss}(T^\infty_{\rm eq})}{2 \,Q \,
    \Omega_B}}
    \approx 1.8 \times 10^{-5}.
    \label{estimate0}
\end{equation}

The time $\Delta t_{\rm torq}$ can also be roughly estimated
if one keeps in mind that,
in the initial stage of the instability,
the amplitude $\alpha$
stays small.
In that case, the first term
in the right-hand side of Eq.\ (\ref{dOmegadt}) can be neglected,
so that one obtains
\begin{equation}
\Omega-\Omega_B \approx \dot{\Omega}_{\rm acc}
(t-t_B).
\label{estimate1}
\end{equation}
Using this equation and expanding into Taylor series the
right-hand side of Eq.\ (\ref{dadt}) around point $B$ (at
fixed $T^\infty=T^\infty_{\rm eq}$), one gets
\begin{equation}
\frac{d \alpha}{dt} \approx \alpha \,\, \frac{|\tau'_{\rm
GR}(\Omega_B)|}{\tau_{\rm GR}^2(\Omega_B)} \,\,
\dot{\Omega}_{\rm acc} \, (t-t_B),
\label{estimate2}
\end{equation}
or, after trivial integration,
\begin{equation}
\alpha = \alpha_B \, {\rm e}^{(t-t_B)^2/\tau_\alpha^2},
\quad \quad {\rm where} \quad \quad
\tau_\alpha=\sqrt{\frac{2 \tau_{\rm
GR}^2(\Omega_B)}{|\tau'_{\rm GR}(\Omega_B)| \,
\dot{\Omega}_{\rm acc}}} \approx 600\, {\rm yr}.
\label{estimate3}
\end{equation}
Substituting now $\alpha=\alpha_0$ into this equation, one finds
\begin{equation}
\Delta t_{\rm torq} \approx \tau_\alpha \,
\sqrt{\ln\left(\frac{\alpha_0}{\alpha_B}\right)} \approx
4600 \,{\rm yr}. \label{estimate4}
\end{equation}
This result is just a little bit larger than the exact
value $\Delta t_{\rm torq}\approx 4500$~yr. As shown in
Sec.\ \ref{Sec_input}, the NS evolution with the saturated
$r^o$-mode is governed by the simpler equations. In
particular, instead of Eq.\ (\ref{dOmegadt}) one will have
\begin{equation}
\frac{d \Omega}{dt} = \frac{2 \, Q \, \alpha_{\rm sat}^2 \,
\Omega}{\tau_{\rm GR}}+\dot{\Omega}_{\rm acc},
\label{dOmegadt1}
\end{equation}
which can be integrated independently.
Neglecting the term
$\dot{\Omega}_{\rm acc}$,
which is small in our case
[$\dot{\Omega}_{\rm acc}$ becomes comparable to the first term
in the right-hand side of Eq.\ (\ref{dOmegadt1}) at a small $\Omega
\approx 915$~s$^{-1}$ ($\nu \approx 146$~Hz),
and can be omitted for a rough estimate],
we get
\begin{equation}
\Omega=\frac{\Omega_B}{(1+t/\tau_\Omega)^{1/6}}, \quad {\rm
with} \quad \tau_\Omega = -\frac{1}{12} \,\,
\frac{\tau_{\rm GR}(\Omega_B)}{Q \, \alpha_{\rm sat}^2}
\approx \frac{3\times 10^{-8}}{\alpha_{\rm sat}^2} \,
\left(\frac{\Omega_0}{\Omega_B}\right)^6 \, {\rm yr}
\approx 7 \times 10^5 \, {\rm yr}, \label{estimate5}
\end{equation}
where the time is counted from the moment when the
$r^o$-mode reaches saturation. In practice, this formula
describes the $\Omega(t)$ dependence on the whole interval
$B$--$C$--$D$ sufficiently well.

Let us now estimate the time $\Delta t_T$ required to heat
up the star from the moment of $r^o$-mode saturation to
point $C$. Point $C$ lies on the Cooling=Heating curve,
given by the condition
\begin{equation}
-\frac{\widetilde{J}  M R^2  \Omega^2  \alpha_{\rm
sat}^2}{\tau_{\rm GR}} -L_{\rm cool} + K_{\rm n} \dot{M}
c^2=0,
\label{Coolheat2}
\end{equation}
which means that at this curve the stellar heating due to
dissipation of the saturated $r^o$-mode is exactly
compensated by the neutrino cooling. After reaching the
curve, the star moves along it, until it enters the
stability region. As we will see from the estimate, $\Delta
t_T$ is much smaller than $\tau_{\Omega}$; thus, in the
further derivation one can set $\Omega=\Omega_B$
($=\Omega_C$) in Eq.\ (\ref{thermal}). Bearing in mind that
the mode is saturated (that is, $\alpha=\alpha_{\rm sat}$
and instead of $\tau_{\rm Diss}$ one should write
$-\tau_{\rm GR}$), Eq.\ (\ref{thermal}) can be rewritten as
\begin{equation}
C_{\rm tot} \frac{dT^\infty}{dt} = -\frac{\widetilde{J}  M
R^2  \Omega_B^2  \alpha_{\rm sat}^2} {\tau_{\rm GR}}
-L_{\rm cool} + K_{\rm n} \dot{M} c^2. \label{thermal1}
\end{equation}
Making the left-hand side of this equation vanish, one
finds the stellar temperature at point $C$, $T^\infty_{C}
\approx 4.6 \times 10^8$~K. Equation (\ref{thermal1}) can
now be integrated in quadratures. However (since we are
only interested in the order-of-magnitude estimate for
$\Delta t_T$), we additionally simplify it by neglecting
the last two terms in the right-hand side of this equation.
In addition, we make use of the fact that in the range of
temperatures under consideration the heat capacity $C_{\rm
tot} \approx \gamma T^\infty$, where $\gamma \approx 1.5
\times 10^{30}$~erg~K$^{-2}$. Integrating now Eq.\
(\ref{thermal1}), we obtain
\begin{equation}
\Delta t_{T} = \frac{\gamma \, \tau_{\rm GR} \, (T_{\rm
eq}^{\infty \, 2}-T_{C}^{\infty \, 2})} {2\widetilde{J}  \,
M \, R^2  \, \Omega_B^2 \,  \alpha_{\rm sat}^2} \approx
1200 \, {\rm yr}. \label{estimate6}
\end{equation}
Because we ignored the star luminosity $L_\mathrm{cool}$ in
Eq.\ (\ref{thermal1}), our rough estimate is smaller than
the real time $\Delta t_T \approx 3300$~yr. In reality,
$L_\mathrm{cool}$ becomes important and slows down the NS
heating only in the very vicinity of the curve
Cooling=Heating. According to our estimate, for the first
$\sim 1300$~yr the star rapidly heats up and reaches the
boundary of the circle, shown as point $C$ in the figure.
During the subsequent $\sim 2000$~yr the star heating
proceeds very slowly and its position in Fig.\
\ref{Fig_typical} almost does not change. As we expected,
$\Delta t_T \ll \tau_{\Omega}$.

($iii$) Stage $C$--$D$.

This is the longest stage in the instability region. During
it the star moves along the Cooling=Heating curve. The time
spent on the horizontal stage $B$--$C$ is several orders of
magnitude smaller. The Cooling=Heating curve crosses the
instability curve at point $D$, $\Omega_D \approx
\Omega_A$. The traveling time along $C$--$D$ can be easily
estimated from Eq.\ (\ref{estimate5}),
\begin{equation}
\Delta t_{CD} \approx \tau_{\Omega} \, \left(
\frac{\Omega_C^6}{\Omega_D^6}-1 \right) \approx \frac{3
\times 10^{-8}}{\alpha_{\rm sat}^2}
\left(\frac{\Omega_0^6}{\Omega_D^6}-\frac{\Omega_0^6}{\Omega_C^6}\right)
\, {\rm yr} \approx 6 \times 10^6 \, {\rm yr}.
\label{estimate7}
\end{equation}
The exact calculation shows that $\Delta t_{CD} \approx 8
\times 10^6$~yr. The discrepancy is due to neglect of the
term $\dot{\Omega}_{\rm acc}$ in the derivation of Eq.\
(\ref{estimate5}).

($iv$) Stage $D$--$A$.

Just after the star reaches the stability region, the
amplitude of $r^o$-mode rapidly (during $\sim 400$~yr)
decreases to negligible values; then the star cools down to
the temperature $T_{\rm eq}^\infty$ (point $A$). The
cooling takes $\sim 10^5$~yr, and after that the cycle
repeats.

The main conclusion that can be drawn from the discussion
of the evolution tracks is as follows: in the {\it
stability} region the star spends most of the time in stage
$A$--$B$, while in the {\it instability} region -- in the
stage $C$--$D$. The ratio of the time spent in the
instability region (without accounting for the time $\Delta
t_{\rm torq}\approx 4500$~yr, during which the star
``sits'' at point $B$; see Fig.\ \ref{Fig_typical}) to the
period of the cycle equals $k\approx0.16$ for the model
with $\alpha_{\rm sat}=10^{-4}$. This ratio drops rapidly
with increasing $\alpha_{\rm sat}$ \cite{heyl02}, because
the typical time $\tau_{\Omega}$ of $\Omega$ variation
during the $C$--$D$ stage is $\tau_{\Omega} \propto
\alpha_{\rm sat}^{-2}$; see Eq.\ (\ref{estimate5}). For
$\alpha_{\rm sat}= 5 \times 10^{-3}$ we have $k \approx
1.7\times10^{-4}$, whereas for $\alpha_{\rm sat}= 10^{-1}$
we obtain $k \approx 10^{-6}$. Let us note that, since in
the saturation regime $W_{\rm Diss}\propto \alpha_{\rm
sat}^2$, the higher $\alpha_{\rm sat}$  is, the farther the
NS gets into the region of high temperatures (the more
horizontally elongated is the track
$A$--$B$--$C$--$D$--$A$; see Fig.\ \ref{Fig_typical}).

\section{NS evolution in the case of
two avoided crossings of oscillation modes}
\label{Appendix_2peaks}

\begin{figure}
    \begin{center}
        \leavevmode
        \includegraphics[width=3.4in]{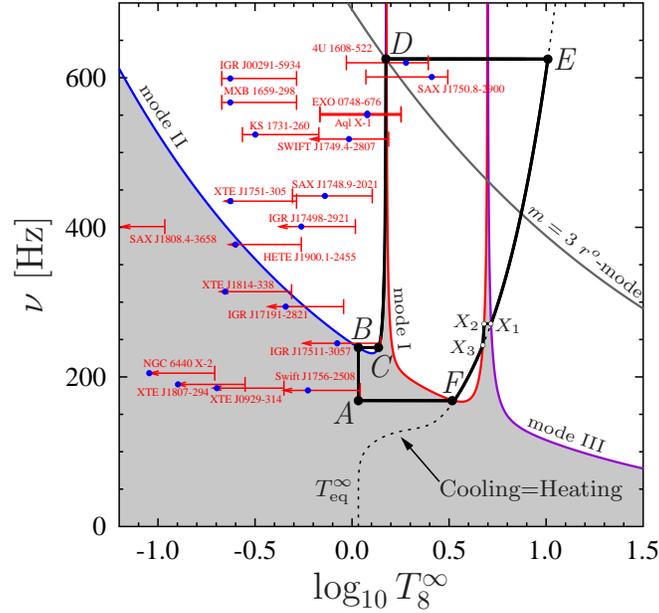}
    \end{center}
    \caption{(color online)
Similar to Fig.\ \ref{Fig_scenario}, but with additional
avoided crossing of modes I and III at
$T^\infty=5\times10^8$~K.
    An instability curve for mode III is shown by the correspondingly marked solid line (violet online);
    the coupling parameter parametrizing
    interaction between modes I and III
    is $s=0.001$.
    The evolution track $A$--$B$--$C$--$D$--$E$--$X_1$--$X_2$--$X_3$--$F$--$A$ of a star
    is shown by the solid line.
    Other notations (and input parameters) coincide with those in Fig. \ref{Fig_scenario}.
    }
    \label{Fig_2peakright}
\end{figure}

Assume that, besides the avoided crossing of modes I and
II, there is one more avoided crossing of modes I and III
at $T^\infty = 5 \times 10^8$~K such that mode III becomes
the $m=2$ $r^o$-mode at $T^\infty>5 \times 10^8$~K (see
Fig. \ref{Fig_2peakright}). The thick solid line in Fig.\
\ref{Fig_2peakright} shows the typical evolution track
$A$--$B$--$C$--$D$--$E$--$X_1$--$X_2$--$X_3$--$F$--$A$ of a
NS in this case. The main difference between this track and
the one discussed in Sec.\ \ref{Sec_climb} (see Fig.\
\ref{Fig_scenario}) is  stage $X_1$--$X_2$--$X_3$, in which
the star evolves in the region
of avoided crossing of modes I and III.%
\footnote{As in Fig.\ \ref{Fig_scenario}, the curve
Cooling=Heating, shown by dots in Fig.\
\ref{Fig_2peakright}, is given by Eq.\ (\ref{Coolheat}) in
the stability region (see also the discussion of this curve
in Sec.\ \ref{Sec_climb}).}

Let us discuss this stage in more detail. At point $X_1$,
the only excited mode is the oscillation mode III, which is
saturated (i.e., its amplitude equals $10^{-4}$). At stage
$X_1$--$X_2$, the star enters the stability region, where
the amplitude of mode III rapidly vanishes, and the star
cools down to point $X_2$ during $\sim130$~yr. At point
$X_2$, the star becomes unstable with respect to excitation
of mode I; similar to the case of mode II at stage $C$--$D$
(see Fig.\ \ref{Fig_2peakright}), its equilibrium amplitude
$\alpha^{\rm (eq)}_{\mathrm I}$ is then defined by the
thermal equilibrium condition (\ref{alpha2}). Since stage
$X_2$--$X_3$ is close to the curve Cooling=Heating,
intensive heating is required to maintain the temperature,
and mode I appears to be close to saturation. Such a high
oscillation amplitude means that the spin-down of the NS
due to viscous dissipation of mode I will dominate the
accretion torque [see\ Eq.\ (\ref{Omega2})]. As a result,
the stellar spin frequency will decrease. Finally, in
$4.6\times10^5$~yr after leaving point $X_2$ (this time
constitutes $\sim 0.16\%$ of the full period of the cycle),
the star again reaches the Cooling=Heating curve at point
$X_3$. To continue spinning down along the instability
curve of mode I, it needs a more intensive heating than the
saturated mode can provide. Thus, the further evolution of
the star (stage $X_3$--$F$) goes along the Cooling=Heating
curve, as in the scenario shown in Fig.\
\ref{Fig_scenario}.

We arrive at the conclusion,
that the existence of additional
avoided crossings of oscillation modes
does not affect noticeably the scenario,
proposed in Sec.\ \ref{Sec_climb},
and does not change the main results of the paper.

\section{NS evolution in the presence of a resonance interaction
of the normal r-mode with the crustal toroidal modes}
\label{Sec_MSP_Levin}

\begin{figure}
    \begin{center}
        \leavevmode
        \includegraphics[width=3.4in]{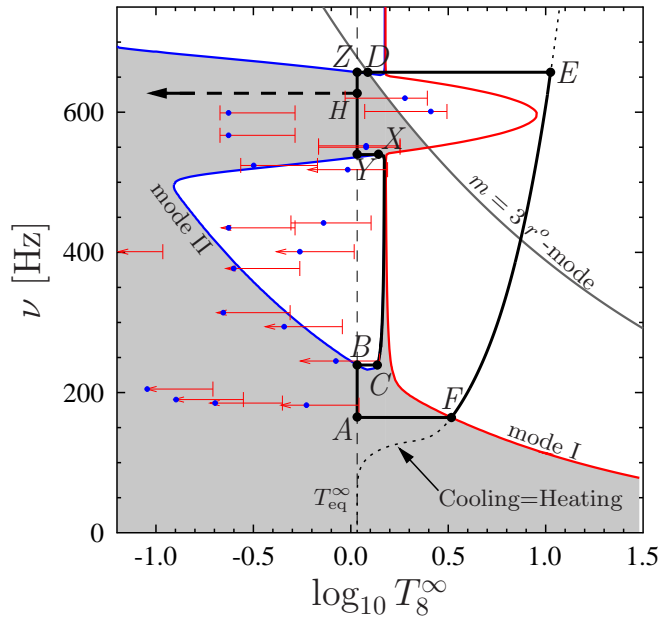}
    \end{center}
    \caption{(color online)
    The same as
    Fig.\ \ref{Fig_scenario},
    but with resonance interaction of the $r$-mode
    and a torsional crustal mode at 600~Hz taken into account.
    The evolution track $A$--$B$--$C$--$X$--$Y$--$Z$--$D$--$E$--$F$--$A$ of a star
    is shown by the thick solid line.
    Horizontal dashes show NS evolution for the case when accretion ceases at point $H$.
    The vertical dashed line indicates $T^\infty_{\rm eq}$.
    }
    \label{Fig_Levin}
\end{figure}

The presence of elastic crust may substantially modify the
oscillation spectrum of rotating NSs. Numerous calculations
(see, e.g., Refs.\ \cite{ls96,lu01,yl01,ga06a}) show that
the $r$-mode in that case experiences avoided crossings
with torsional crustal modes at some spin frequencies. An
important feature of $r$-mode eigenfunctions near these
frequencies is amplified relative velocity (slippage)
between the elastic crust and liquid core. This
amplification leads to an enhanced damping in the Ekman
layer \cite{lu01} near the avoided crossing that could
modify the instability windows \cite{lu01,hah11}. In Fig.\
\ref{Fig_Levin} we show a possible example of such a
modified instability window. For illustration, we assume
that there is only one avoided crossing of the normal
$r$-mode and a superfluid inertial mode (see Fig.\
\ref{Fig_scenario}). In addition, we assume that the normal
($m=2$) $r$-mode experiences a resonant interaction with
the torsional crustal mode at a rotation frequency
$\nu_{\rm crust}=600\,\rm Hz$ \cite{lu01,hah11}. The
resulting enhanced dissipation in the Ekman layer is
modeled, in a simplified manner, by introducing an
additional frequency-dependent term in the expression for
the total inverse damping time scale $1/\tau_{\rm
Diss}^{\rm norm}$ of the normal $r$-mode,
\begin{equation}
\frac{1}{\tau_{\rm Ek}^{\rm norm}}= \frac{0.08 \, \rm s^{-1}}{(T^\infty_8)^2}\, {\rm exp} \left[-1133\left(\frac{\nu-\nu_{\rm crust}}{1 \,\rm kHz}\right)^2\right],
\label{tauEk}
\end{equation}
so that now $1/\tau_{\rm Diss}^{\rm norm} =1/\tau_{\rm
S}^{\rm norm}+1/\tau_{\rm MF}^{\rm norm}+1/\tau_{\rm
Ek}^{\rm norm}$ (while the corresponding inverse damping
time scale $1/\tau_{\rm Diss}^{\rm sfl}$ for the superfluid
mode is kept unchanged). The functional dependence and
numerical values in Eq.\ (\ref{tauEk}) are purely
illustrative. This additional dissipation leads to the
appearance of the stability region at spin frequencies
close to $\nu_{\rm crust}$ (see the filled grey region in
Fig.\ \ref{Fig_Levin}) and modifies the evolutionary track
of a NS, shown by the thick solid line.
Stages $A$--$B$--$C$ are the same as in Fig.\
\ref{Fig_scenario}. From point $C$ a star climbs up the
left edge of the stability peak until it reaches  point
$X$, where ongoing accretion brings it inside the stability
region. The next stage $X$--$Y$ is similar to  stage
$F$--$A$: In the stability region, the $r$-mode dies out
and the star cools down to $T_{\rm eq}^\infty$ (point $Y$).
Then it spins up slowly in the stability region (like in
the $A$--$B$ stage). At point $Z$, the star becomes
unstable with respect to the $m=2$ $r$-mode and starts to
heat up rapidly. At point $D$ it becomes, in addition,
unstable with respect to the $m=3$ $r$-mode. The subsequent
stages $D$--$E$--$F$--$A$ are analogous to those shown in
Fig.\ \ref{Fig_scenario}. Note that, if accretion ceases at
the stage $Y$--$Z$ (e.g., at point $H$), then the star will
cool down (see the thick arrowed dashed line) and can
become a very cold MSP with $\nu \ga 500$~Hz.

\end{document}